# What's in a game?

## A theory of game models

Clovis Eberhart
Univ. Savoie Mont Blanc, CNRS, LAMA
Chambéry, France
clovis.eberhart@univ-smb.fr

Tom Hirschowitz
Univ. Savoie Mont Blanc, CNRS, LAMA
Chambéry, France
tom.hirschowitz@univ-smb.fr

**Abstract**

Game semantics is a rich and successful class of denotational models for programming languages. Most game models feature a rather intuitive setup, yet surprisingly difficult proofs of such basic results as associativity of composition of strategies. We set out to unify these models into a basic abstract framework for game semantics, *game settings*. Our main contribution is the generic construction, for any game setting, of a category of games and strategies. Furthermore, we extend the framework to deal with innocence, and prove that innocent strategies form a subcategory. We finally show that our constructions cover many concrete cases, mainly among the early models [6, 20] and the very recent sheaf-based ones [37].

## 1 Introduction

Game semantics has provided adequate models for a variety of (idealised) programming languages. We will here mainly be concerned with the numerous variations on *arena games*. This comprises, e.g., the original dialogue game model of PCF [20, 33], Abramsky et al.'s model for general references [5], Harmer and McCusker's model for finite nondeterminism [15, 17], Laird's model for control operators [24], and the recent model by Tsukada and Ong [37]. We will also briefly consider other models of PCF [6] and of linear logic [10].

This rich literature shares many features. E.g., all these models follow the same simple conceptual route: the types of the considered language are interpreted as games and programs as strategies. Games form the objects, and strategies the morphisms of a category, which is compared to the 'syntactic' category generated from the operational description of the language. However, as noted, e.g., in Harmer et al. [16], a less advantageous feature shared by all models is the surprising difficulty of certain proofs like associativity of composition or the fact that innocent strategies are closed under composition.

This raises the issue of unifying all these models into a satisfactory theory, with an emphasis on factoring out such difficult proofs. This is an ambitious goal, because although game models clearly share a lot of ideas, they are also rather diverse. E.g., depending on the considered language, various constraints are imposed upon strategies, like *innocence* or *well-bracketing*. Further sources of diversity have appeared with recent extensions, e.g., sheaf-based innocence [37], nominal models [32], tensorial logic [30], or concurrent extensions [18, 34].

This paper is an attempt at improving the situation. Focussing on the construction of game models, our main contributions are:

- We organise the basic data underlying typical game models into a simple categorical structure called a *game setting*, emphasising its simplicial features. Each game setting gives rise to a candidate category of games and strategies.
- We delineate mild hypotheses under which games and strategies do form a category, heavily relying on high-level categorical techniques, including (presheaf) polynomial functors [23, Chapter 16] and Guitart's *exact squares* [14].
- We extend the framework to deal with *innocence*, an emblematic constraint put on strategies to capture purely functional computation. We enrich game settings with a notion of *view* and, under mild hypotheses, we derive a category of innocent strategies. Our approach exploits the recent recasting of innocence as a sheaf condition [18, 37], and again relies on advanced category theory to give high-level proofs.
- We show that a number of game models fall into our framework, namely variants of the original Hyland-Ong/Nickau (HON) model [15, 20, 26], AJM games [6] and Tsukada and Ong's model [37].
- We work out the limits of our techniques in two well-known dead ends of game semantics: non-associativity of composition in Blass games [4] and non-stability of innocent strategies under composition in the absence of determinism [15].
- In passing, we prove a categorical result of possibly independent interest, stating that certain commuting squares of categories and functors, called *local pushforward squares*, are *distributive*, in the sense that doing left then right (Kan) extension along one side is isomorphic to doing right then left along the other.

Our framework deals with various notions of composition and innocence. For clarity, let us readily fix some terminology. An important distinction is whether plays are considered as a poset (with the prefix ordering) or as a category. Another is whether the considered strategies are general or boolean presheaves. We annotate composition and innocence with the following codes.

| Plays | | Strategies | |
|---|---|---|---|
| poset | category | boolean | general |
| p | c | b | s |

**Table 1.** Conventions for plays and strategies

**Example 1.1.** Standard strategies, being prefix-closed sets of plays, are boolean presheaves on the prefix ordering, so their composition is pb-composition. Moreover, standard innocent strategies are innocent pb-strategies. Similarly, Tsukada and Ong [37] use proper categories of plays and their strategies are presheaves, hence their composition and innocent strategies are cs-composition and innocent cs-strategies. Unspecified items denote either possibility. E.g., s-composition means composition of presheaves, in either the poset-based or category-based setting.

Much effort is put into linking the different variants together, as summarised in the following table.

| Section | relates | to |
|---|---|---|
| 2.8 | s-composition | b-composition |
| 4.2 | cs-innocence | ps-innocence |
| 4.3 | cs-innocence | pb-innocence |

The established links are of various nature. E.g., in Section 2.8, we both infer associativity of b-composition from associativity of s-composition, and express b-composition in terms of s-composition.

### 1.1 Related work

Although significant work has been devoted to giving efficient proofs in particular models [16, 28, 30], the only attempts at abstraction we are aware of are our Hirschowitz [18] and Eberhart and Hirschowitz [12]. Both papers focus on the link between naive and innocent strategies as well as the interpretation of programs, but both fail to capture composition of strategies.

### 1.2 Plan

In Section 2, we gradually introduce game settings, following the successive steps for constructing a typical game model. We also state our main results for the basic setup along the way. We remain very informal about game semantics, and only start to consider the particulars of various game models in Section 3, where we establish that the announced game models fit into our framework. We then refine game settings to deal with innocence in Section 4, covering in passing Tsukada and Ong's model [37]. Finally, we conclude in Section 5. Most proofs are deferred to appendices.

### 1.3 Notation and prerequisites

For all $n \in \mathbb{N}$, $[n]$ denotes the finite ordinal with $n$ elements, i.e., the set $\{0, \ldots, n-1\}$, and we sometimes use just $n$ to denote the set $\{1, \ldots, n\}$.

We assume some basic knowledge of category theory, namely categories, functors and natural transformations, as well as adjunctions. The category of *presheaves* over any category $\mathcal{C}$ is the functor category $[\mathcal{C}^{op}, \text{Set}]$ of contravariant functors to sets and natural transformations between them, which we denote by $\widehat{\mathcal{C}}$. For any presheaf $X \colon \mathcal{C}^{op} \to \text{Set}$, objects $c, c' \in \mathcal{C}$, morphism $f \colon c' \to c$, and element $x \in X(c)$, we use a right action notation $x \cdot f$ for $X(f)(x)$. By functoriality, we have $x \cdot f \cdot g = x \cdot (f \circ g)$, for any $g \colon c'' \to c'$. Replacing Set with 2, the ordinal 2 viewed as a category, we get the category $\widetilde{\mathcal{C}}$ of *boolean presheaves*.

## 2 Game settings

### 2.1 Categories of plays in game semantics

In this section, we sketch several notions of play typically involved in the construction of a game model. We do this without referring to any particular model. In the next sections, we will organise this data into a coherent categorical structure, which we will then exploit to give an abstract construction of a category of games and strategies.

The construction of a typical game model relies on the definition of increasingly complex notions of play. There is first a notion of *game*. Each game $A$ involves two players $O$ (Opponent) and $P$ (Proponent), and features in particular a set of *plays* $\mathbb{P}_A$, which may be endowed with the prefix ordering or with a more sophisticated notion of morphism, thus forming a category of plays. Such two-player games form the basis of the model.

The crucial step to view strategies as morphisms is to consider the *arrow* game $A \to B$, which intuitively describes the interaction of a middle player $M$ playing Opponent against a left player $L$ and Proponent against a right player $R$, as in

$$
\begin{array}{ccc}
L & M & R \\
\mathbb{B} & \longrightarrow \mathbb{B} & \\
& & q^R \\
q^M & & \\
\mathsf{t}^L & & \\
& \mathsf{f}^M. &
\end{array} \quad (1)
$$

In this example, $M$ plays like the negation function on booleans: $R$ asks its return value by playing the move $q^R$; $M$ in turn asks $L$ for the value of the argument by playing $q^M$, to which $L$ answers 'true' by playing $\mathsf{t}^L$; $M$ eventually answers the original question by playing $\mathsf{f}^M$.

However, there is a subtlety: one often needs to restrict plays in $\mathbb{P}_{A \to B}$ with additional constraints, so that the relevant category is a subcategory $\mathbb{P}_{A,B} \hookrightarrow \mathbb{P}_{A \to B}$. Of course, there are projections to $\mathbb{P}_A$ and $\mathbb{P}_B$.

**Example 2.1.** We will provide more precise definitions later on, but for now, to fix intuition, in HON-style games (without bracketing) $\mathbb{P}_A$ would consist of all justified sequences, and $\mathbb{P}_{A,B}$ would restrict to alternating sequences of even length of $\mathbb{P}_{A \to B}$. The projections of a play in $\mathbb{P}_{A,B}$ to $A$ and $B$ may not be alternating, so it is crucial to be liberal in the choice of $\mathbb{P}_A$.

In order to define composition of strategies, the situation (1) is then scaled up to combinations of two such situations in which a first middle player $M_1$ plays on the right with a second one, say $M_2$, as in

$$
\begin{array}{cccc}
L & M_1 & M_2 & R \\
\mathbb{B} & \longrightarrow \mathbb{B} & \longrightarrow \mathbb{B}.
\end{array} \quad (2)
$$

Plays in such combinations are standardly called *interaction sequences*, and typically form a subcategory $\mathbb{P}_{A,B,C} \hookrightarrow \mathbb{P}_{(A \to B) \to C}$. An important point is that interaction sequences admit projections to $\mathbb{P}_{A,B}$, $\mathbb{P}_{B,C}$ and $\mathbb{P}_{A,C}$, which satisfy the obvious equations w.r.t. further projections to $\mathbb{P}_A$, $\mathbb{P}_B$ and $\mathbb{P}_C$, e.g., the following square commutes:

$$
\begin{array}{ccc}
\mathbb{P}_{A,B,C} & \longrightarrow & \mathbb{P}_{A,C} \\
\downarrow & & \downarrow \\
\mathbb{P}_{B,C} & \longrightarrow & \mathbb{P}_C.
\end{array}
$$

**Example 2.2.** In HON games, $\mathbb{P}_{A,B,C}$ typically consists of alternating justified sequences on $(A \to B) \to C$ which end in $A$ or $C$ and whose projections to $A \to B$ and $B \to C$ are plays.

Finally, in order to prove associativity of composition, one defines *generalised interaction sequences* as a subcategory $\mathbb{P}_{A,B,C,D} \hookrightarrow \mathbb{P}_{((A \to B) \to C) \to D}$, again with projections satisfying the obvious equations.

## 2.2 Plays as a category-valued presheaf

Let us now organise all this data ($\mathbb{P}_A, \mathbb{P}_{A,B}, \mathbb{P}_{A,B,C}, \mathbb{P}_{A,B,C,D}$) into a simple categorical structure. First, as suggested by our notation, for all lists $L = (A_0, \ldots, A_{n-1})$ of games, we may construct a category $\mathbb{P}_L$.

**Example 2.3.** In the HON case, we may take $\mathbb{P}_L$ to consist of alternating justified sequences on $(\ldots (A_0 \to A_2) \to \ldots) \to A_{n-1}$ whose projection to each $A_i \to A_{i+1}$ is a play, and which end in $A_0$ or $A_{n-1}$, the rightmost arena.

For the same reasons as before, we get projections $\delta_k \colon \mathbb{P}_L \to \mathbb{P}_{L \setminus A_k}$ (for 'delete $k$'), for all $k \in [n]$. A similar construction, relevant for defining identity strategies (so-called *copycat* strategies), is *insertions* $\iota_k \colon \mathbb{P}_L \to \mathbb{P}_{L^{+k}}$ (for 'insert $k$'), where $k \in [n]$ and $L^{+k}$ denotes $L$ with the $k$th game duplicated. E.g., $\iota_1 \colon \mathbb{P}_{A,B} \to \mathbb{P}_{A,B,B}$. Intuitively, this functor maps any play $u$ in $\mathbb{P}_{A,B}$ to the interaction sequence in $\mathbb{P}_{A,B,B}$ which duplicates all moves on $B$. So in a situation like (2), $M_2$ would act as a 'proxy' between $M_1$ and $R$, repeating $M_1$'s moves to $R$ and conversely. For a precise definition and an example in the case of HON games, see Section 3.1.

This may all be packed up into the comma category $\Delta/\mathbb{A}$, or more precisely $i/\ulcorner \mathbb{A} \urcorner$, where

- $\ulcorner \mathbb{A} \urcorner \colon 1 \to \mathrm{Set}$ is the functor picking $\mathbb{A}$;
- $i \colon \Delta \hookrightarrow \mathrm{Set}$ is the embedding of the simplicial category $\Delta$ into sets.

Let us recall that $\Delta$ has finite ordinals $[n]$ as objects, with monotone maps as morphisms. So concretely, $\Delta/\mathbb{A}$ has finite lists of games as objects, i.e., maps $L \colon [n] \to \mathbb{A}$ for some $n = \{0, \ldots, n-1\}$, and as morphisms $(n, L) \to (n', L')$ all monotone maps $f$ making the following triangle commute:

$$\begin{array}{ccc} [n] & \xrightarrow{f} & [n'] \\ {}_L \searrow & & \swarrow {}_{L'} \\ & \mathbb{A}. & \end{array}$$

**Example 2.4.** Let $\mathrm{d}_k^n \colon [n] \to [n+1]$ miss $k \in [n]$, i.e., $\mathrm{d}_k^n(i) = i$ for $i < k$ and $\mathrm{d}_k^n(j) = j+1$ for $j \geq k$. Then, e.g., $\mathrm{d}_1^2$ yields a map $(A, C) \to (A, B, C)$ for all games $A, B, C$. Similarly, consider $\mathrm{i}_k^n \colon [n+1] \to [n]$ which collapses $k \in [n] \subseteq [n+1]$ and $k+1 \in [n+1]$. E.g., for $n = 2$ and $k = 0$, it yields a map $(A, A, B) \to (A, B)$ for all $A$ and $B$.

As promised, this yields a way to organise the various categories of plays involved in a typical game model into a coherent categorical structure. Indeed, we will show below that, for quite a few game models, the assignment $L \mapsto \mathbb{P}_L$ induces a category-valued presheaf on $\Delta/\mathbb{A}$, i.e., a functor $(\Delta/\mathbb{A})^{op} \to \mathrm{Cat}$. Furthermore, the maps $\delta_k$ and $\iota_k$ introduced earlier will respectively be given by $\mathbb{P}(\mathrm{d}_k)$ and $\mathbb{P}(\mathrm{i}_k)$.

In the following, we will only need to use this structure up to lists of length 4:

**Definition 2.5.** For any $p \leq q$ and set $\mathbb{A}$, let $\mathbb{A}_{[p,q]}$ denote the full subcategory of $\Delta/\mathbb{A}$ spanning lists $L$ of length between $p$ and $q$.

In the next sections, we will define strategies, composition and copycat strategies abstractly, based on the category-valued presheaf $\mathbb{P}$ on $\mathbb{A}_{[1,4]}$. This is quite demanding, but we are rewarded with a higher-level view of composition, which yields abstract proofs of associativity and unitality, assuming a few additional properties of $\mathbb{P}$. We will define a game setting to consist of a set $\mathbb{A}$ and a category-valued presheaf satisfying these additional properties.

## 2.3 Notions of strategy

Let us now start our reconstruction of a game model from an arbitrary category-valued presheaf $\mathbb{P}$ on $\mathbb{A}_{[1,4]}$. Our first step is to define strategies. Standardly, a strategy $\sigma \colon A \to B$ is a prefix-closed set of plays in $\mathbb{P}_{A,B}$ (generally required to be non-empty). Equivalently, it is a functor $\mathbb{P}_{A,B}^{op} \to 2$, the ordinal 2 viewed as a category. In Tsukada and Ong's model [37], $\mathbb{P}_{A,B}$ is a proper category, and strategies are generalised to *presheaves* on $\mathbb{P}_{A,B}$, i.e., functors $\mathbb{P}_{A,B}^{op} \to \mathrm{Set}$. This is indeed a generalisation because 2 embeds into Set (more on this in Section 2.8).

The basis of our approach will be the general notion:

**Definition 2.6.** Let the category of *strategies* from $A$ to $B$ be $\widehat{\mathbb{P}_{A,B}}$. The category of *boolean strategies* is $\widetilde{\mathbb{P}_{A,B}}$.

## 2.4 Polynomial functors

The next step in our reconstruction of a game model from $\mathbb{A}$ and $\mathbb{P}$ is to define identities and composition, which will rely on polynomial functors, which we now briefly recall.

**Notation 1.** *Any functor* $F \colon \mathcal{C} \to \mathcal{D}$ *induces a restriction, or pre-composition functor* $\Delta_F \colon \widehat{\mathcal{D}} \to \widehat{\mathcal{C}}$ *mapping any* $X \colon \mathcal{D}^{op} \to \mathrm{Set}$ *to* $X \circ F^{op}$, *where* $F^{op} \colon \mathcal{C}^{op} \to \mathcal{D}^{op}$ *acts just as $F$ but on opposite categories. When $\mathcal{C}$ and $\mathcal{D}$ are small, this restriction functor has both a left and a right adjoint, which we respectively denote by $\sum_F$ and $\prod_F$, as in*

$$\widehat{\mathcal{C}} \xleftarrow{\underset{\prod_F}{\overset{\sum_F}{\rightleftarrows}}} \widehat{\mathcal{D}}.$$

The left and right adjoints are respectively given by left and right extension, and enjoy explicit descriptions, both in terms of *coends* and *ends* and in terms of colimits and limits [25, 35]. A brief description of how they work is included for completeness in Appendix A.1, but most of the paper should be accessible without reading it.

**Definition 2.7.** A functor is *polynomial* iff it is isomorphic to some finite composite of functors of the form $\Delta_F$, $\prod_F$ and $\sum_F$.

This definition is very close to Kock [23, Chapter 16] – it might even turn out to be the same if Kock's presheaf polynomial functors are ever proved to be closed under composition.

## 2.5 Copycat as a polynomial functor

As a warm-up before considering composition, we would like to start with our abstract definition of identities, which are standardly given by *copycat* strategies. A natural way to define the copycat strategy $id_A \colon A \to A$ is to decree that it accepts all plays in $\mathbb{P}_{A,A}$ which are in the image of the insertion functor $\iota_0 \colon \mathbb{P}_A \to \mathbb{P}_{A,A}$. Indeed, recalling (1) and according to the discussion of insertions, right after Example 2.3, such plays are precisely those in which $M$ acts as a proxy between $L$ and $R$, which agrees with the standard definition of copycat strategies.

This definition has the advantage of concreteness, but as announced we need to give an equivalent, polynomial definition. Because an object of a category $\mathbb{C}$ is the same as a functor $1 \to \mathbb{C}$, we may define $id_A$ as a functor $1 \to \widehat{\mathbb{P}_{A,A}}$. Furthermore, 1 is a presheaf category: indeed it is $\widehat{\emptyset}$, presheaves over the empty category. So we may view copycat over $A$ as a functor $\widehat{\emptyset} \to \widehat{\mathbb{P}_{A,A}}$. In order to present it as a polynomial functor, we will need to assume that the insertion functor $\iota_0 \colon \mathbb{P}_A \to \mathbb{P}_{A,A}$ is a discrete fibration. Let us recall the definition:

**Definition 2.8.** A functor $p \colon \mathcal{E} \to \mathcal{B}$ is a *discrete fibration* when for all objects $e \in \mathcal{E}$ and morphisms $f \colon b \to p(e)$ there exists a unique morphism $u \colon e' \to e$ such that $p(u) = f$. Such a morphism is called a *cartesian lifting* of $e$ along $f$. Let $\mathrm{DFib}_{\mathcal{B}}$ denote the full subcategory of $\mathrm{Cat}/\mathcal{B}$ spanning discrete fibrations.

For us, the relevant property of discrete fibrations is the following characterisation of left extension along them:

**Lemma 2.9.** *For any discrete fibration $p \colon \mathcal{E} \to \mathcal{B}$, presheaf $X \in \widehat{\mathcal{E}}$, and object $b \in \mathcal{B}$, we have $\sum_p(X)(b) \cong \sum_{e | p(e) = b} X(e)$, where $\sum$ means left extension in the left-hand side and disjoint union on the right-hand one.*

Here is our polynomial presentation of copycat:

**Proposition 2.10.** *If the insertion functor $\iota_0 \colon \mathbb{P}_A \to \mathbb{P}_{A,A}$ is a discrete fibration, then the functor*

$$\widehat{\emptyset} \xrightarrow{\prod_!} \widehat{\mathbb{P}_A} \xrightarrow{\sum_{\iota_0}} \widehat{\mathbb{P}_{A,A}}$$

*is isomorphic the copycat strategy $id_A$.*

*Proof.* See Section B.1. □

### 2.6 Composition as a polynomial functor

The next step is to express composition of strategies using the same language of polynomial functors. Let us first recall the standard definition in the boolean case: the composite $\sigma; \tau$ of two boolean strategies $\sigma$ and $\tau$ over $(A, B)$ and $(B, C)$ respectively, is defined to accept all plays $p \in \mathbb{P}_{A,C}$ for which there exists $u \in \mathbb{P}_{A,B,C}$ such that $\delta_1(u) = p$ and

$$\delta_2(u) \in \sigma \qquad \text{and} \qquad \delta_0(u) \in \tau.$$

In Tsukada and Ong [37], this is extended to a polynomial functor $\widehat{\mathbb{P}_{A,B}} \times \widehat{\mathbb{P}_{B,C}} \to \widehat{\mathbb{P}_{A,C}}$, whose definition is essentially a proof-relevant version of the boolean one. If we understand '$\sigma$ accepts play $p$' as $\sigma(p) = \{\star\}$, then, e.g., $\delta_2(u) \in \sigma$ above becomes $\star \in \sigma(\delta_2(u))$. We get:

**Definition 2.11.** The composite $\sigma; \tau$ of two strategies $\sigma$ and $\tau$ over $(A, B)$ and $(B, C)$ respectively, is defined to map any play $p \in \mathbb{P}_{A,C}$ to the set of triples $(u, x, y)$ where $u \in \mathbb{P}_{A,B,C}$ is such that $\delta_1(u) = p$ and

$$x \in \sigma(\delta_2(u)) \qquad \text{and} \qquad y \in \tau(\delta_0(u)).$$

Let us present this polynomially. First, by universal property of coproduct we have $\widehat{\mathbb{P}_{A,B}} \times \widehat{\mathbb{P}_{B,C}} \cong \widehat{\mathbb{P}_{A,B} + \mathbb{P}_{A,B}}$, so we reduce to defining a functor $\widehat{\mathbb{P}_{A,B} + \mathbb{P}_{B,C}} \to \widehat{\mathbb{P}_{A,C}}$. Here is our candidate composition:

**Definition 2.12.** Let m denote the polynomial functor

$$\widehat{\mathbb{P}_{A,B} + \mathbb{P}_{B,C}} \xrightarrow{\Delta_{\delta_2 + \delta_0}} \widehat{\mathbb{P}_{A,B,C} + \mathbb{P}_{A,B,C}} \xrightarrow{\prod_{[id, id]}} \widehat{\mathbb{P}_{A,B,C}} \xrightarrow{\sum_{\delta_1}} \widehat{\mathbb{P}_{A,C}}.$$

This definition is legitimated by:

**Proposition 2.13.** *The functor m agrees with Definition 2.11, i.e., for all $\sigma$ and $\tau$, we have $(\sigma; \tau) \cong \mathrm{m}[\sigma, \tau]$.*

*Proof sketch (see Section B.1 for a full proof).* By discrete fibredness, the final $\sum_{\delta_1}$ says that for all $p \in \mathbb{P}_{A,C}$, the formula for $\mathrm{m}[\sigma, \tau](p)$ will start by

$$\sum_{u \in \mathbb{P}_{A,B,C} | \delta_1(u) = p} \ldots (u).$$

By a very general computation, the $\prod_{[id, id]}$ says that, viewing the intermediate result in $\widehat{\mathbb{P}_{A,B,C} + \mathbb{P}_{A,B,C}}$ as a pair $[\sigma', \tau']$, the formula continues with $\sigma'(u) \times \tau'(u)$, which in our case directly instantiates to $\sigma(\delta_2(u)) \times \tau(\delta_0(u))$, as desired. □

**Remark 1.** *The discrete fibredness hypothesis is satisfied in most game models, with the notable exception of the saturated interpretation of AJM games (see Section 3.4), in which the projection is a non-discrete fibration. The construction still goes through, but we here stick to discrete fibrations for simplicity.*

### 2.7 Game settings, associativity and unitality

We have now expressed copycat strategies and composition abstractly, relying only on the postulated category-valued presheaf. Let us now consider associativity. It has become standard in game semantics to prove associativity of composition using a *zipping* result [8] stating that both squares

$$\begin{array}{ccc} \mathbb{P}_{A,B,C,D} \longrightarrow \mathbb{P}_{A,B,D} & \mathbb{P}_{A,B,C,D} \longrightarrow \mathbb{P}_{A,C,D} \\ \downarrow \quad \lrcorner \quad \downarrow & \downarrow \quad \lrcorner \quad \downarrow \\ \mathbb{P}_{B,C,D} \longrightarrow \mathbb{P}_{B,D} & \mathbb{P}_{A,B,C} \longrightarrow \mathbb{P}_{A,C} \end{array} \quad (3)$$

are pullbacks on objects. This holds in all considered game models, and constitutes the last bit of our axiomatisation:

**Definition 2.14.** A *game setting* consists of a set $\mathbb{A}$ (whose elements we call *arenas*) and a category-valued presheaf $\mathbb{P}$ on $\mathbb{A}_{[1,4]}$ such that all projections $\mathbb{P}_{A,B,C} \to \mathbb{P}_{A,C}$ and insertions $\mathbb{P}_A \to \mathbb{P}_{A,A}$ are discrete fibrations, and all squares (3) are pullbacks on objects.

We call both squares (3) the *zipping* squares of $\mathbb{P}$.
One of our main results is:

**Theorem 2.15.** *In any game setting, composition of strategies is associative up to isomorphism, and copycat strategies are units up to isomorphism.*

*Proof.* See Section B.2. □

### 2.8 The boolean case

Let us conclude this section by treating the boolean case: until now, our strategies were given by general presheaves (Definition 2.6). We would like to derive from Theorem 2.15 that boolean strategies also form a category.

The bridge to the boolean case is given by the embedding $r \colon 2 \hookrightarrow$ Set mapping $0 \leq 1$ to $\emptyset \to 1$. This functor has a left adjoint l mapping $\emptyset$ to 0 and collapsing all non-empty sets to 1. Furthermore, r being fully faithful, we have in fact a full reflection (please note: this means that the right adjoint is fully faithful, not the left adjoint!), which induces a further one between presheaves and boolean presheaves:

What's in a game?

**Proposition 2.16.** *For any small category $\mathbb{C}$, post-composition by $\mathsf{l}$ and $\mathsf{r}$ yield a full reflection*

$$[\mathbb{C}^{op}, \mathsf{Set}] \xrightleftharpoons[\mathsf{r}_!]{\mathsf{l}_!} \perp \; [\mathbb{C}^{op}, 2].$$

*The left adjoint $\mathsf{l}_!$ is called* booleanisation.

*Proof.* The functor $\mathsf{r}_!$ is clearly a continuous full embedding, hence a right adjoint by the special adjoint functor theorem, and thus a full, reflective embedding. □

Because 2 is complete and cocomplete, replacing Set with 2 in Notation 1 yields a notion of *boolean* polynomial functor.

**Notation 2.** *Any functor $F\colon \mathcal{C} \to \mathcal{D}$ induces restriction, left extension and right extension functors between boolean presheaf categories $\widetilde{\mathcal{C}}$ and $\widetilde{\mathcal{D}}$, respectively denoted by $\overline{\Delta_F}$, $\overline{\Sigma_F}$ and $\overline{\prod_F}$. Accordingly, the boolean version of any polynomial functor $P$ will be denoted by $\overline{P}$.*

We may thus transfer our polynomial definitions of copycat and composition to boolean strategies. Concrete examples of game settings will be considered in Section 3, for which we have:

**Proposition 2.17.** *In all the game settings of Section 3, $\overline{m}$ coincides with standard composition.*

As desired, we obtain:

**Proposition 2.18.** *In any game setting, composition of boolean strategies is associative and unital up to isomorphism.*

*Proof.* See Section B.4. □

**Remark 2.** *Please note that we have not claimed that boolean composition agrees with general, set-based composition, i.e., commutation of the left-hand diagram below.*

$$\begin{array}{ccc} \widetilde{\mathbb{P}_{A,B} + \mathbb{P}_{B,C}} & \xrightarrow{m} & \widetilde{\mathbb{P}_{A,C}} \\ {\scriptstyle \mathsf{r}_!}\uparrow & & \uparrow{\scriptstyle \mathsf{r}_!} \\ \widehat{\mathbb{P}_{A,B} + \mathbb{P}_{B,C}} & \xrightarrow[\overline{m}]{} & \widehat{\mathbb{P}_{A,C}} \end{array}$$

*In fact it does not in general, and this is the main cause for the failure of stability of boolean, innocent strategies under composition [15, Section 3.7.2]. What does hold, however, is*

- *commutation of booleanisation with composition as on the left below (this is the main idea for the proof of Proposition 2.18),*
- *the characterisation of boolean composition given below right, as set-based composition followed by booleanisation.*

$$\begin{array}{ccc} \widehat{\mathbb{P}_{A,B} + \mathbb{P}_{B,C}} & \xrightarrow{m} & \widehat{\mathbb{P}_{A,C}} \\ {\scriptstyle \mathsf{l}_!}\downarrow & & \downarrow{\scriptstyle \mathsf{l}_!} \\ \widetilde{\mathbb{P}_{A,B} + \mathbb{P}_{B,C}} & \xrightarrow[\overline{m}]{} & \widetilde{\mathbb{P}_{A,C}} \end{array} \qquad \begin{array}{ccc} \widehat{\mathbb{P}_{A,B} + \mathbb{P}_{B,C}} & \xrightarrow{m} & \widehat{\mathbb{P}_{A,C}} \\ {\scriptstyle \mathsf{r}_!}\uparrow & & \downarrow{\scriptstyle \mathsf{l}_!} \\ \widetilde{\mathbb{P}_{A,B} + \mathbb{P}_{B,C}} & \xrightarrow[\overline{m}]{} & \widetilde{\mathbb{P}_{A,C}} \end{array}$$

Let us move on to exhibit a few concrete game settings. We will return to the boolean case in Section 4.3 where we consider boolean innocence.

## 3 Applications

In this section, we show that a number of standard game models fit into our framework. In Section 3.1, we consider HON games, in their p-form first, which by the results of Section 2 yields categories of ps and pb-strategies. We then refine our results by considering variants in which some constraints are imposed on strategies (or equivalently plays): a first, local form of constraint is treated in Section 3.2; a more global form of constraint, obtained by enriching games with validity predicates on plays, is considered in Section 3.3. All these variants are shown to form game settings (hence yield categories of ps and pb-strategies). AJM games are considered in Section 3.4, and also shown to form a game setting. Finally, we explain in Section 3.5 why Blass games fail to form a game setting.

### 3.1 Hyland-Ong games and strategies

Let us now consider HON games in more detail, and show that they form a game setting. We mostly follow Harmer's [15] presentation. For simplicity, we adopt the following innocuous[1] modification of the standard notion of arena:

**Definition 3.1.** An *arena* is a simple, countable, directed acyclic graph $A$ equipped with a subset $\sqrt{A}$ of *initial* vertices, or *roots*, such that for all vertices $m$, all paths from $m$ to some initial vertex have the same parity.

In particular, simple, upside-down forests form arenas. The intuition is that reachable vertices of an arena are moves in a two-player game, and that an edge $m \to m'$ in the forest means that $m$ is *enabled*, or *justified* by $m'$. If the path from $m$ to some root has even length, then $O$ (for *Opponent*) is playing; otherwise $P$ (for *Proponent*) is. E.g., all roots are played by $O$.

**Example 3.2.** A very simple arena, called $o$, is the single-vertex graph. For a less trivial example, the boolean type $\mathbb{B}$ may be interpreted as the arena 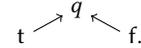

Now that we have defined arenas, let us move on to define plays. The idea, explained at length, e.g., in McCusker [27], is that plays are sequences of moves in which $O$ and $P$ take turns. But a subtlety is that moves may be played several times. So for any edge $m \to m'$ in the considered arena, there may be several occurrences of $m$ and $m'$. We thus decorate sequences of moves with justification pointers matching those of the considered arena.

**Definition 3.3.** A *justified sequence* on any arena $A$ consists of a natural number $n \in \mathbb{N}$, equipped with maps $f\colon n \to \mathsf{ob}(A)$ and $\varphi\colon n \to \{0\} \uplus n$ (recalling Section 1.3) such that, for all $i \in n$,

- $\varphi(i) < i$,
- if $\varphi(i) = 0$ then $f(i) \in \sqrt{A}$, and
- if $\varphi(i) \neq 0$, then there is an edge $f(i) \to f(\varphi(i))$ in $A$.

Let $\mathbb{P}_A$ denote the poset of justified sequences on $A$, with prefix ordering.

We will draw justified sequences $(n, f, \varphi)$ as the sequence of their $f(i)$'s, with arrows to denote $\varphi$, as is standard in game semantics.

**Example 3.4.** Here is a justified sequence in the boolean arena: 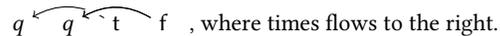, where times flows to the right.

As mentioned in (1), game semantics proceeds by letting a middle player $M$ play on two arenas $A$ and $B$, with specific restrictions on when switching between $A$ and $B$ is allowed. For this, we form the compound arena $A \to B$:

---
[1]Tsukada and Ong use forest-shaped arenas.

**Definition 3.5.** For any two arenas $A$ and $B$, let $A \to B$ denote the arena obtained by taking the disjoint union of $A$ and $B$ as directed graphs, adding an edge $m \to m'$ for all $m \in \sqrt{A}$ and $m' \in \sqrt{B}$, and taking $\sqrt{B}$ as $\sqrt{A \to B}$ (if $B$ is not empty, otherwise we take $A \to B$ to be empty).

The switching constraints mentioned above are implemented by considering a subposet of $\mathbb{P}_{A \to B}$:

**Definition 3.6.** For any two arenas $A$ and $B$, let $\mathbb{P}_{A,B}$ denote the poset of *plays* on $(A, B)$, i.e., *alternating* justified sequences of even length on $A \to B$.

Alternation here means that, for any $s = (n, f, \varphi)$, $f(i)$ is played by $O$ iff $i$ is odd.

**Example 3.7.** Recalling the arena $o$ from Example 3.2, and calling its unique move $r$, here is an example play on $((\mathbb{B} \to o), o)$:

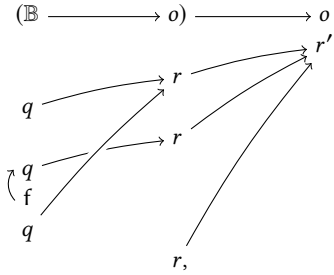

where time flows downwards (so the play really is $r'rqrqfqr$) and arrows denote justification pointers.

We are now in a position to define the insertion functors $\iota_0 \colon \mathbb{P}_A \to \mathbb{P}_{A,A}$: they send any justified sequence $m_1 \ldots m_n$ to some play in which $P$ mimics the behaviour of $O$. The technical definition is not particularly illuminating, but the example following it should convince the reader that nothing really difficult is going on here.

**Definition 3.8.** For any justified sequence $p = (m_1 \ldots m_n)$ on $A$, let $\iota_0(p)$ be the sequence $(m_1, k_1)(m_1, l_1) \ldots (m_n, k_n)(m_n, l_n)$, where $k_i$ and $l_i$ denote either 0 or 1, according to the component of $A \to A$ in which the move is played. If $m_i$ is an $O$-move, then $k_i = 1$ and $l_i = 0$; and otherwise $k_i = 0$ and $l_i = 1$. Pointers are as in $p$ except for initial moves on the left, i.e., moves of the form $(m, 0)$ with $m \in \sqrt{A}$, which are justified by the corresponding $(m, 1)$.

**Example 3.9.** The justified sequence on $\mathbb{B}$ below left, which is not alternating, yields the copycat play on the right:

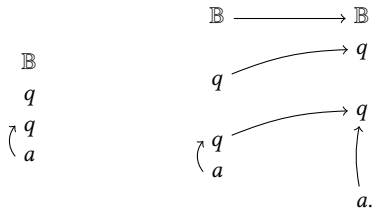

The next step is interaction sequences, for which the basic idea is: any play in $(A \to B) \to C$ may be projected to $\mathbb{P}_{A \to B}$, $\mathbb{P}_{B \to C}$, and even $\mathbb{P}_{A \to C}$, by prolongating pointers (i.e., $a \to b \to c$ becomes $a \to c$). Following Example 2.2, we put:

**Definition 3.10.** An *interaction sequence* is a justified sequence on $(A \to B) \to C$ ending in $A$ or $C$, whose projections to $A \to B$ and $B \to C$ are plays. Let $\mathbb{P}_{A,B,C}$ denote the poset of interaction sequences with prefix ordering.

As desired, the projection to $A \to C$ is also a play, and we have monotone maps $\delta_k \colon \mathbb{P}_{A_0, A_1, A_2} \to \mathbb{P}_{A_i, A_j}$ with $i < j$ in $\{0, 1, 2\} \setminus \{k\}$.

We may define *generalised interaction sequences* similarly to obtain:

**Proposition 3.11.** *The category-valued presheaf $\mathbb{P}$ defined by respectively taking $\mathbb{P}_A$, $\mathbb{P}_{A,B}$, $\mathbb{P}_{A,B,C}$ and $\mathbb{P}_{A,B,C,D}$ to be the posets of all justified sequences, plays, interaction sequences and generalised interaction sequences, for all arenas $A, B, C, D$, with projections and insertions as above, forms a game setting.*

*Proof.* Projections $\delta_1 \colon \mathbb{P}_{A,B,C} \to \mathbb{P}_{A,C}$ are discrete fibrations: the restriction of any $u \in \mathbb{P}_{A,B,C}$ along any $p \leq \delta_1(u)$ may be taken to be the shortest prefix of $u$ whose projection is $p$ (longer such prefixes do not end in $A$ or $C$). The fact that squares (3) are pullbacks on objects is the standard zipping lemma. □

### 3.2 Constraining strategies: local constraints

In the previous section, we consider a rather rough notion of play. Standardly, further constraints are considered on strategies, such as *P*-visibility, *O*-visibility, well-threadedness, and well-bracketing (when games are equipped with an appropriate question-answer discipline). E.g., a *P*-visible strategy is one which only accepts *P*-visible plays. One then needs to prove that such constraints are *robust*, i.e., are preserved by composition and satisfied by identities. This is done in a very clean and modular way in Harmer's thesis [15, Chapter 3].

In order for our framework to apply to such constrained strategies, we may start from the game setting for unconstrained plays and convert the proof of robustness of constraint $c$ into the construction of a sub-game setting $\mathbb{P}^c$.

For each constraint $c \in \{P\text{-}vis, O\text{-}vis, wb, wt\}$, respectively denoting *P*-visibility, *O*-visibility, well-bracketing and well-threadedness:

- $\mathbb{P}^c_{A,B}$ is the full sub-poset of $\mathbb{P}_{A,B}$ consisting of plays satisfying the constraint $c$;
- $\mathbb{P}^c_A$ is the full sub-poset of $\mathbb{P}_A$ whose insertions are in $\mathbb{P}^c_{A,A}$;
- $\mathbb{P}^c_{A,B,C}$ is the full sub-poset of $\mathbb{P}_{A,B,C}$ whose projections to $\mathbb{P}_{A,B}$ and $\mathbb{P}_{B,C}$ respectively factor through $\mathbb{P}^c_{A,B}$ and $\mathbb{P}^c_{B,C}$;
- $\mathbb{P}^c_{A,B,C,D}$ is the full sub-poset of $\mathbb{P}_{A,B,C,D}$ whose projections to $\mathbb{P}_{A,B}$, $\mathbb{P}_{B,C}$ and $\mathbb{P}_{C,D}$ respectively factor through $\mathbb{P}^c_{A,B}$, $\mathbb{P}^c_{B,C}$ and $\mathbb{P}^c_{C,D}$.

One delicate point is then to show that unmentioned projections also factor through the appropriate constrained posets.

**Proposition 3.12.** *For all constraints $c$, $\mathbb{P}^c_{A,B,C} \to \mathbb{P}_{A,C}$ factors through $\mathbb{P}^c_{A,C}$.*

*Proof.* For *P*-visibility, this is [15, Proposition 3.4.3]. The (implicit) proof of [26, Lemma 2.8] handles *O*-visibility and well-bracketing. Well-threadedness follows similarly to *P*-visibility. □

**Proposition 3.13.** *For all constraints $c$, $\mathbb{P}^c$ forms a sub-game setting of $\mathbb{P}$.*

*Proof.* First, to see that projections $\delta_1$ are discrete fibrations, it is enough to observe that all involved constraints are stable under prefix. It remains to show that constrained plays satisfy zipping. But constraints are merely imposed on the projections of interaction sequences, and thus are clearly stable under zipping. □

**Corollary 3.14.** *For all sets of constraints in $\{P\text{-vis}, O\text{-vis}, wb, wt\}$, arenas and strategies satisfying these constraints form a category.*

As satisfactory as it seems, this result does not tell us that composition is the considered sub-game settings agrees with the original. Let us show that it does, by considering the general case. Consider any embedding $c \colon \mathbb{P}^c \hookrightarrow \mathbb{P}$ of game settings sharing a common set of arenas. We need to prove that the polynomial functor, say $\mathsf{m}^c$, mimicking our polynomial composition $\mathsf{m}$ on $\mathbb{P}^c$ is compatible with the inclusion $c \colon \mathbb{P}^c \hookrightarrow \mathbb{P}$. In order to make this more precise, let us observe that each strategy $\sigma \in \widehat{\mathbb{P}^c_{A,B}}$ is canonically mapped to $\sum_c(\sigma) \in \widehat{\mathbb{P}_{A,B}}$. Intuitively, $\sum_c(\sigma)$ should act just as $\sigma$ on $\mathbb{P}^c_{A,B}$ and be empty elsewhere. In order for this to hold, we merely need to assume that the considered constraint is stable under prefix, which in the general case amounts to requiring $c$ to be a discrete fibration.

**Lemma 3.15.** *If $c$ is a discrete fibration, then for any $\sigma$ we have*

- $\sum_c(\sigma)(c(p)) \cong \sigma(p)$ *for all $p \in \mathbb{P}^c_{A,B}$ and*
- $\sum_c(\sigma)(q) = \emptyset$ *for all $q \in \mathbb{P}_{A,B}$ outside the essential image of $c$.*

*Proof.* By Lemma 2.9. □

We may now express the desired compatibility of composition with $c$ as $\mathsf{m}(\sum_c(\sigma), \sum_c(\tau)) \cong \sum_c(\mathsf{m}^c(\sigma, \tau))$. This however requires an additional hypothesis, saying that an interaction sequence $u \in \mathbb{P}_{A,B,C}$ is in $\mathbb{P}^c_{A,B,C}$ as soon as its projections to $\mathbb{P}_{A,B}$ and $\mathbb{P}_{B,C}$ are.

**Example 3.16.** To see why, imagine that there exists some interaction sequence $u \in \mathbb{P}_{A,B,C} \setminus \mathbb{P}^c_{A,B,C}$ such that $\delta_2(u)$ and $\delta_0(u)$ are in the respective essential images of $\mathbb{P}^c_{A,B}$ and $\mathbb{P}^c_{B,C}$, say as $c(p_1)$ and $c(p_2)$. Further assuming that $\sigma$ and $\tau$ accept $p_1$ and $p_2$, $\mathsf{m}(\sum_c(\sigma), \sum_c(\tau))$ clearly accepts $\delta_1(u)$, while $\sum_c(\mathsf{m}^c(\sigma, \tau))$ does not, because the interaction sequence which could witness it lies outside $\mathbb{P}^c$ (assuming that no interaction sequence from $\mathbb{P}^c$ projects to $\delta_1(u)$).

This hypothesis essentially says that the considered constraint may be checked locally, hence our terminology:

**Definition 3.17.** *An embedding $c \colon \mathbb{P}^c \hookrightarrow \mathbb{P}$ between game settings sharing the same set of arenas is a local constraint iff its components are discrete fibrations and any $u \in \mathbb{P}_{A,B,C}$ is essentially in the image of $c$ if $\delta_2(u)$ and $\delta_0(u)$ are.*

**Remark 3.** *The locality constraint may be expressed concisely as a sheaf condition (see Lemma B.9).*

**Proposition 3.18.** *For all local constraints $c \colon \mathbb{P}^c \hookrightarrow \mathbb{P}$, the square*

$$\begin{array}{ccc} \widehat{\mathbb{P}^c_{A,B}} \times \widehat{\mathbb{P}^c_{B,C}} & \xrightarrow{\mathsf{m}^c} & \widehat{\mathbb{P}^c_{A,C}} \\ {\scriptstyle \sum_c \times \sum_c} \downarrow & & \downarrow {\scriptstyle \sum_c} \\ \widehat{\mathbb{P}_{A,B}} \times \widehat{\mathbb{P}_{B,C}} & \xrightarrow{\mathsf{m}} & \widehat{\mathbb{P}_{A,C}} \end{array}$$

*commutes up to isomorphism.*

*Proof.* See Section B.5. □

### 3.3 Constraining strategies: predicates

Beyond the constraints mentioned above, a similar result may be proved for the refined notion of game in McCusker's thesis [26]. McCusker's games $A$ are just like arenas, except that they come equipped with an abstract *validity* predicate $P_A$, which is a subset of the set $L_A$ of *legal plays*, i.e., alternated, well-bracketed, $P$- and $O$-visible justified sequences. This predicate is only required to be non-empty, prefix-closed, and such that for all $p \in P_A$ and set $I$ of occurrences of initial moves in $p$, the restriction $p_{|I}$ of $p$ to moves hereditarily justified by some move in $I$ is again in $P_A$.

McCusker then defines $P_{A \to B}$ to consist of legal plays in $L_{A \to B}$ whose projections to $A$ and $B$ are in $P_A$ and $P_B$ (instead of simply $L_A$ and $L_B$), respectively. Interaction sequences are then defined as in the standard case, with the additional constraint that both projections to $(A, B)$ and $(B, C)$ are in $P_{A \to B}$ and $P_{B \to C}$, respectively.

Let us now show that this again yields a game setting. Denoting the new posets by $\mathbb{P}^P_A, \mathbb{P}^P_{A,B}$, etc., we have:

**Proposition 3.19.** *Projections $\mathbb{P}^P_{A,B,C} \to \mathbb{P}_{A,C}$ factors through $\mathbb{P}^P_{A,C}$.*

*Proof.* Being in $P_{A \to C}$ is only about projections being in $P_A$ and $P_C$, which is taken care of by the condition on projections to $(A, B)$ and $(B, C)$. □

Similarly:

**Proposition 3.20.** $\mathbb{P}^P$ *is a game setting, and a local constraint of $\mathbb{P}$.*

*Proof.* As before, projections are discrete fibrations because validity predicates are stable under prefix. It only remains to show that zipping holds, which again follows from $P_{A \to B}$ being only about projections to $A$ and $B$. □

There are other kinds of constraints like innocence or single-threadedness, which may not be treated this way. We will deal with innocence in Section 4.

### 3.4 AJM games: a partial answer

Let us now briefly consider an alternative approach to game semantics by Abramsky et al. [6]. On the one hand, this approach is more elementary than Hyland and Ong's in that games do not feature justification pointers. So, e.g., composition of strategies is significantly simpler. On the other hand, games feature a partial equivalence relation between plays, which needs to be dealt with at the level of strategies.

In order to organise such games into a game setting, we have two sensible choices for the notion of morphism between plays: beyond the prefix ordering, we may also incorporate equivalence between plays. Presheaves then amount to so-called *saturated* strategies. We adopt Harmer's presentation [15].

**Definition 3.21.** *A game $A$ consists of two sets $O_A$ and $P_A$, respectively of Opponent and Proponent moves, equipped with a partial, prefix-closed equivalence relation $\approx$ on alternated sequences of moves started by Opponent, for which any two equivalent plays have the same length, and such that*

*if $s \approx t$ and $sa \approx sa$, then there exists $a'$ such that $sa \approx ta'$.*

Let $\mathbb{P}_A$ consist of all alternated sequences of moves $s$ started by Opponent, such that $s \approx s$. Then, for any games $A$ and $B$, we form the game $A \to B$, which has $O_{A \to B} = P_A + O_B$ and $P_{A \to B} = O_A + P_B$ and $s \approx_{A \to B} t$ iff $s$ and $t$ play in the same component at each stage and their projections are equivalent in $A$, resp. $B$. The poset $\mathbb{P}_{A,B}$ may then be defined as the set of plays in $A \to B$ equipped with prefix ordering. Similarly defining $\mathbb{P}_{A,B,C}$ and $\mathbb{P}_{A,B,C,D}$ we obtain:

**Proposition 3.22.** *AJM games form a game setting.*

*Proof.* Squares (3) being pullbacks on objects is the standard zipping lemma. To show that projections $\delta_1 \colon \mathbb{P}_{A,B,C} \to \mathbb{P}_{A,C}$ are discrete fibrations, we need to be able to canonically restrict any $u \in \mathbb{P}_{A,B,C}$ along any $s \leq \delta_1(u)$: just take the longest prefix of $u$ mapped to $s$. □

For saturated strategies, the idea is to incorporate for all $A$, $B$ the partial equivalence relations $\approx_A$ and $\approx_B$ into the category of plays.

**Proposition 3.23.** *AJM games form a category-valued presheaf by mapping each list of games to the corresponding set of plays with as morphisms between any two plays $u$ and $v$:*

- *a singleton when there exists some play $w$ such that $u \approx w \leq v$, or equivalently there exists $w$ such that $u \leq w \approx v$;*
- *none otherwise.*

However, the obtained category-valued presheaf is not a game setting, because projections $\mathbb{P}_{A,B,C} \to \mathbb{P}_{A,C}$ and insertions $\mathbb{P}_A \to \mathbb{P}_{A,A}$ are not discrete fibrations. Indeed, the fibres of $\mathbb{P}_{A,B,C} \to \mathbb{P}_{A,C}$ are proper groupoids in general, thus making it a non-discrete Grothendieck fibration. The case of $\mathbb{P}_A \to \mathbb{P}_{A,A}$ is worse: the restriction of a play in $s \in \mathbb{P}_A$ along a morphism $p \to q$ in $\mathbb{P}_{A,A}$ may at best be mapped to some $p'$ isomorphic to $p$ in general, thus making it a *Street fibration*. Our approach may generalise in this direction, but this will involve advanced categorical concepts such as stacks (which are to fibrations as sheaves are to discrete fibrations), so we leave it for future work.

### 3.5 A non-example: Blass games

In the previous sections, we have shown that several approaches to game semantics form game settings, with the exception of the saturated AJM setting. It may be instructive to consider Blass's games [9, 10], as they are well-known for their non-associative composition. Our account essentially follows Abramsky [4, Section 3], through the lens of game settings.

**Definition 3.24.** A Blass game consists of a family of rooted trees, together with a *polarity* in $\{P, O\}$.

Vertices are thought of as positions in the game, with alternating polarities. The given polarity indicates which player is to start the game, by choosing the initial position. The fact that Proponent may start is a notable difference with arena games. Another difference is that the given family of trees genuinely represents the 'game tree' – no move may be played twice. This determines the definition of $\mathbb{P}_A$, for any game $A = (T, \pi)$: it is the poset consisting of positions (i.e., vertices of $T$, plus a formal initial position), with $x \leq y$ when $x$ is above $y$ in $T$.

For $\mathbb{P}_{A,B}$ things become a bit more complicated. Strategies should be based on the *linear implication* game $A \to B$, which is constructed much as in, e.g., AJM games. First, one lets $A^\perp$ denote the game with the same family of trees as $A$ but with opposite polarity. Then, one defines $A \to B$ by interleaving moves from $A^\perp$ and $B$ with natural switching conditions: Opponent is to play as soon as possible. In other words, if the respective polarities in $A^\perp$ and $B$ are $OP$, $PO$, or $OO$, then $O$ is to play; otherwise $P$ is. There is a catch, however: if the polarity is $OO$ and Opponent plays, say in $B$, we reach a position with polarity $OP$, and Opponent is to play again, which breaks alternation. This is rectified by defining $A \to B$ to comprise *compound* moves from $OO$ to $PP$, for each pair of moves in $A^\perp$ and $B$. This settles the definition of $\mathbb{P}_{A,B}$, up to some technicalities.

The next level is to define $\mathbb{P}_{A,B,C}$. Glossing over the details, this should consist of sequences of moves in $A$, $B$ and $C$, whose projections to $\mathbb{P}_{A,B}$, $\mathbb{P}_{B,C}$ and $\mathbb{P}_{A,C}$ are well-defined. However, we may show that with these definitions, the squares (3) cannot both be pullbacks in general. Indeed, consider the case where the respective polarities of $A$, $B$, $C$ and $D$ are $O$, $P$, $O$ and $P$, and $A$ is non-empty. Then, let $\mathbb{P}^l_{A,B,C,D}$ denote the left-hand pullback and $\mathbb{P}^r_{A,B,C,D}$ denote the right-hand one. We will show that both pullbacks cannot be the same category by exhibiting a play in $\mathbb{P}^l_{A,B,C,D}$ which is not in $\mathbb{P}^r_{A,B,C,D}$. First, let us observe that the initial polarities from the respective points of view of $A \to B$, $B \to C$ and $C \to D$ are like so:

$$
\begin{array}{cccccc}
A & \longrightarrow & B & B & \longrightarrow & C & C & \longrightarrow & D \\
A^\perp & & B & B^\perp & & C & C^\perp & & D \\
P & & P & O & & O & P & & P.
\end{array}
$$

Letting $a$ denote any root of $A$, the sequence $a$ is then legal in $\mathbb{P}_{A,B,D}$ (the polarities are $PP$ both in $A \to B$ and $A \to D$) and the empty sequence is legal in $\mathbb{P}_{B,C,D}$. Thus, $a$ is legal in $\mathbb{P}^l_{A,B,C,D}$ by the left-hand pullback. However, if the two pullbacks were isomorphic, then by the properties of projections $a \in \mathbb{P}^l_{A,B,C,D}$ would be mapped to $a \in \mathbb{P}^r_{A,B,C,D}$ under the isomorphism. But $\mathbb{P}^r_{A,B,C,D}$ cannot contain $a$ because this play is illegal in $\mathbb{P}_{A,B,C}$ (because the polarity is $PO$ in $A \to C$).

## 4 Innocence
### 4.1 Concurrent innocence

In the previous sections, we have constructed a category of games and strategies parameterised over an arbitrary game setting which presents the advantage of unifying a number of such categories as instances of the same construction. However, in game models of purely functional languages, the relevant category is the identity-on-objects subcategory of innocent strategies. In this section, we extend game settings with a notion of view, which allows us to construct a subcategory of innocent strategies.

In order to achieve this, we will use the recent recasting of innocence as a sheaf condition [18, 19, 37]. Starting from HON games, the first step is to refine the posets $\mathbb{P}_A, \mathbb{P}_{A,B}, \ldots$ into proper categories (with exactly the same objects), say $\mathbb{P}^+_A, \mathbb{P}^+_{A,B}, \ldots$, with the crucial feature that for any play $p \in \mathbb{P}^+_{A,B}$ and move $m \in p$, there is a morphism $\lceil p \rceil_m \to p$ from the $P$-view of $m$ to $p$[2]. This of course does not hold with the prefix ordering, as the view is rarely a prefix. This idea was introduced by Melliès [28] in a slightly different setting.

Passing from $\mathbb{P}$ to $\mathbb{P}^+$ raises the issue of how to extend the abstract framework. Should it now contain two category-valued presheaves? Or should we simply forget about prefix-based strategies and accept $\mathbb{P}^+$ as the new basic set up? We do not make any definitive choice here, but for simplicity and modularity reasons, we choose to first

---

[2]We omit the definition of views, as it is unnecessary for understanding the rest.

work with $\mathbb{P}^+$ only, and in a second round explore the connection with $\mathbb{P}$.

Indeed, perhaps surprisingly, we have:

**Proposition 4.1.** *Tsukada and Ong's $\mathbb{P}^+$ forms a game setting.*

*Proof.* Follows from Lemmas 39, 46 and 47 of Tsukada and Ong [36]. □

Returning to the abstract setting, the new data thus merely consists of a full subcategory $i_{A,B} : \mathbb{V}_{A,B} \hookrightarrow \mathbb{P}^+_{A,B}$, for all $A, B$, whose objects are called *views*.

**Definition 4.2.** The category of *innocent* strategies is the essential image of $\prod_{i_{A,B}} : \widehat{\mathbb{V}_{A,B}} \to \widehat{\mathbb{P}^+_{A,B}}$. The domain $\widehat{\mathbb{V}_{A,B}}$ is the category of *behaviours*.

We now would like to establish that in any game setting equipped with such full embeddings, innocent strategies form a subcategory. However, our proof relies on two additional properties. The first, already observed by Tsukada and Ong [37, Lemma 32], states that one can reconstruct uniquely any interaction sequence from its projection to $\mathbb{P}_{A,C}$, say $u$, together with a compatible family, for each view $v$ of $u$ of an interaction sequence projecting to $v$. The second property essentially says that any morphism $v \to \delta_2(u)$ from a view $v \in \mathbb{V}_{A,B}$ to the projection of some $u \in \mathbb{P}^+_{A,B,C}$ factors canonically through the projection of some view (and similarly for $\delta_0$).

Let us introduce introduce both properties in more detail.

The first property essentially says that interaction is local. To state it, we need to recall the following standard construction and its prime property.

**Definition 4.3.** The *category of elements* $\mathbb{C}/X$ of any presheaf $X \in \widehat{\mathbb{C}}$ has as objects all pairs $(c, x)$ with $x \in X(c)$ and as morphisms $(c, x) \to (c', x')$ all morphisms $f : c \to c'$ such that $x' \cdot f = x$. Let $p_X : \mathbb{C}/X \to \mathbb{C}$ map any $(c, x)$ to $c$, and any $f$ to itself.

**Proposition 4.4.** *The assignment $X \mapsto (\mathbb{C}/X, p_X)$ extends to a functor* el $: \widehat{\mathbb{C}} \to \mathrm{DFib}_\mathbb{C}$ *(recalling Definition 2.8).*

*Proof.* The cartesian lifting of $(c', x')$ along any $f : c \to c'$ is given by $f$ itself, viewed as a morphism $(c, x' \cdot f) \to (c', x')$ in $\mathbb{C}/X$. □

This functor el is in fact an adjoint equivalence.

**Definition 4.5.** Consider the functor sing $: \mathrm{DFib}_\mathbb{C} \to \widehat{\mathbb{C}}$ mapping any discrete fibration $p : \mathbb{E} \to \mathbb{C}$ to the presheaf $\mathrm{sing}(p)(c) = p^{-1}(c)$, with action on morphisms given by cartesian lifting.

**Proposition 4.6.** *The functors*

$$\widehat{\mathbb{C}} \xrightleftharpoons[\mathrm{sing}]{\mathrm{el}} \mathrm{DFib}_\mathbb{C}$$

*form an adjoint equivalence.*

Let us now state the first property we need to impose on game settings with embeddings $i_{A,B} : \mathbb{V}_{A,B} \hookrightarrow \mathbb{P}^+_{A,B}$. The projection $\mathbb{P}_{A,B,C} \to \mathbb{P}_{A,C}$, as a discrete fibration, induces a presheaf $\mathrm{sing}(\mathbb{P}_{A,B,C})$ on $\mathbb{P}_{A,C}$ which we will require to be in the essential image of

$$\prod_{i_{A,B}} : \widehat{\mathbb{V}_{A,C}} \to \widehat{\mathbb{P}_{A,C}}.$$

This is equivalent to requiring that $\mathrm{sing}(\mathbb{P}_{A,B,C})$ be a *sheaf* for the Grothendieck topology induced by the embedding $\mathbb{V}_{A,C} \to \mathbb{P}_{A,C}$. Similarly, we require the presheaf induced by $\iota_0 : \mathbb{P}_A \to \mathbb{P}_{A,A}$ to be a sheaf for the Grothendieck topology induced by the embedding $\mathbb{V}_{A,A} \to \mathbb{P}_{A,A}$. Let us record this as:

**Definition 4.7.** A game setting $(\mathbb{A}, \mathbb{P})$ equipped with full embeddings $i_{A,B} : \mathbb{V}_{A,B} \hookrightarrow \mathbb{P}_{A,B}$ is *local* iff $\mathrm{sing}(\mathbb{P}_{A,B,C})$ and $\mathrm{sing}(\mathbb{P}_A)$ are sheaves.

**Proposition 4.8.** *Tsukada and Ong's $\mathbb{P}^+$ is local.*

*Proof.* For $\mathbb{P}_{A,B,C}$, the result is precisely [37, Lemma 32]. For $\mathbb{P}_A$, just observe that a play is copycat iff all its views are. □

So locality is the first property we need to require of our game settings with views. The second property has to do with projections, e.g., $\delta_2 : \mathbb{P}_{A,B,C} \to \mathbb{P}_{A,B}$. It essentially says that any morphism $v \to \delta_2(u)$ with $v \in \mathbb{V}_{A,B}$ and $u \in \mathbb{P}_{A,B,C}$ factors 'canonically' through some $\delta_2(w)$ with $w \in \mathbb{V}_{A,B,C}$, where $\mathbb{V}_{A,B,C}$ denotes the full subcategory of $\mathbb{P}_{A,B,C}$ projecting to $\mathbb{V}_{A,C}$ (or otherwise said, $\mathbb{V}_{A,B,C} = \mathbb{P}_{A,B,C} \times_{\mathbb{P}_{A,C}} \mathbb{V}_{A,C}$). In order to define such canonicity, we appeal to the theory of analytic functors [21, 38, 39].

**Definition 4.9** (Weber [38, 39]). A functor $T : \mathbb{C} \to \mathbb{D}$ admits *generic factorisations* relative to an object $d \in \mathbb{D}$ iff any $f : d \to Tc$ admits a factorisation as below left

$$\begin{array}{ccc} d & \xrightarrow{f} & \\ g \downarrow & & \\ Ta & \xrightarrow{Th} & Tc \end{array} \qquad \begin{array}{ccc} d & \xrightarrow{g'} & Tb \\ g \downarrow & \overset{Tk}{\nearrow} & \downarrow Th' \\ Ta & \xrightarrow{Th} & Tc \end{array}$$

such that for all commuting squares as the exterior above right, there exists a lifting $k$ as shown making the diagram commute, or more precisely such that $g' = Tk \circ g$ and $h = h'k$. The middle object $a$ is called the *arity* of $f$ – all such factorisations share the same $a$ up to isomorphism.

For all subcategories $\mathbb{B} \hookrightarrow \mathbb{C}$ and $\mathbb{E} \hookrightarrow \mathbb{D}$, a functor $\mathbb{C} \to \mathbb{D}$ admitting generic factorisations relative to all objects of $\mathbb{E}$ with arities in $\mathbb{B}$ is called $(\mathbb{B}, \mathbb{E})$-*analytic* [13].

**Definition 4.10.** A game setting $(\mathbb{A}, \mathbb{P})$ equipped with full embeddings $i_{A,B} : \mathbb{V}_{A,B} \hookrightarrow \mathbb{P}_{A,B}$ is *view-analytic* when $\delta_2$ is $(\mathbb{V}_{A,B,C}, \mathbb{V}_{A,B})$-analytic and $\delta_0$ is $(\mathbb{V}_{A,B,C}, \mathbb{V}_{B,C})$-analytic.

**Proposition 4.11.** *Tsukada and Ong's game setting [37] is view-analytic.*

*Proof.* This follows from [36, Lemma 36]. □

We may now state our main result about innocence:

**Definition 4.12.** An *innocent game setting* is a game setting $(\mathbb{A}, \mathbb{P})$ equipped with full embeddings $i_{A,B} : \mathbb{V}_{A,B} \hookrightarrow \mathbb{P}_{A,B}$, which is both local and view-analytic.

**Theorem 4.13.** *In any innocent game setting, innocent strategies form a subcategory.*

Again, we defer the proof to the more technical Section B.6.

## 4.2 Prefix-based innocence

In the previous section, we have shown that innocent strategies behave well in any innocent game setting. However, our only concrete example of an innocent game setting for now is Tsukada and Ong's $\mathbb{P}^+$. There is in fact a further example, given by enriching arenas with bracketing information and restricting $\mathbb{P}^+_{A,B}$ to well-bracketed plays [37, Section VII]. This shows that innocence is stable under cs-composition. How about pb-composition? As mentioned before, innocence is not stable under pb-composition in general [15, Section 3.7.2], unless one restricts to deterministic strategies. In an attempt to better understand this phenomenon, we first move in this section from cs-composition to ps-composition, and prove that innocence remains stable. In the next section, we will explain why this does not carry over to pb-composition, although, as is well-known, it does on deterministic strategies.

We here proceed by first defining innocent, prefix-based strategies in an extended framework and then showing that our definition agrees with the standard one (which is only defined on boolean behaviours). We then show that ps-innocent strategies include copycats and are closed under composition.

**Definition 4.14.** Consider game settings $(\mathbb{A}, \mathbb{P}^+)$ and $(\mathbb{A}, \mathbb{P})$ with the same set of arenas and $\mathbb{V}$ making $\mathbb{P}^+$ innocent, further equipped with a componentwise identity-on-objects natural embedding $k \colon \mathbb{P} \hookrightarrow \mathbb{P}^+$. Let a presheaf on $\mathbb{P}_{A,B}$ be *innocent via* $\mathbb{P}^+$, or $\mathbb{P}^+$-*innocent*, iff it is in the essential image of $\widehat{\mathbb{V}_{A,B}} \xrightarrow{\Pi_{i_{A,B}}} \widehat{\mathbb{P}^+_{A,B}} \xrightarrow{\Delta_{k_{A,B}}} \widehat{\mathbb{P}_{A,B}}$. Similarly, let a presheaf on $\mathbb{P}_{A,A}$ be $\mathbb{P}^+$-*copycat* iff it is in the essential image of $1 \cong \widehat{\emptyset} \xrightarrow{\Pi_!} \widehat{\mathbb{P}^+_A} \xrightarrow{\Sigma_{\iota_0}} \widehat{\mathbb{P}^+_{A,A}} \xrightarrow{\Delta_{k_{A,A}}} \widehat{\mathbb{P}_{A,A}}$.

**Proposition 4.15.** *A strategy in the standard HON sense is innocent iff it is $\mathbb{P}^+$-innocent. It is copycat iff it is $\mathbb{P}^+$-copycat.*

*Proof.* We prove the more specific statement that the innocent strategy associated to any boolean behaviour (i.e., in this case, a behaviour in the image of $rr_!$) in standard HON games is given up to isomorphism by the functor $\Delta_{k_{A,B}} \circ \Pi_{i_{A,B}}$. Similarly, the copycat on $A$ is given by the image of the unique element of $1$ under $\Delta_{k_{A,A}} \circ \Sigma_{\iota_0} \circ \Pi_!$. For innocence, consider any boolean behaviour $B$. As is standard, right extension may in this case be computed as a conjunction, which entails that $B$ is mapped to

$$B'(p) = \bigwedge_{\{v \in \mathbb{V}_{A,B} | \mathbb{P}^+_{A,B}(v,p) \neq \emptyset\}} B(v),$$

i.e., $p$ is accepted iff $B$ accepts all its views. The case of copycats is similar. $\square$

By the proposition, we may understand ps-innocence through cs-innocence. Let us now state the transfer result.

**Proposition 4.16.** *In the setting of Definition 4.14, if all naturality squares*

$$\begin{array}{ccc} \mathbb{P}_{A,B,C} & \longrightarrow & \mathbb{P}_{A,C} \\ \downarrow & & \downarrow \\ \mathbb{P}^+_{A,B,C} & \longrightarrow & \mathbb{P}^+_{A,C} \end{array} \quad \text{and} \quad \begin{array}{ccc} \mathbb{P}_A & \longrightarrow & \mathbb{P}_{A,A} \\ \downarrow & & \downarrow \\ \mathbb{P}^+_A & \longrightarrow & \mathbb{P}^+_{A,A} \end{array}$$

*are pullbacks, then $\mathbb{P}^+$-innocent strategies are closed under composition and comprise $\mathbb{P}^+$-copycat strategies.*

The proof is in Section B.7. Of course, both hypotheses are satisfied in the case of HON games.

## 4.3 Boolean innocence

We finally consider boolean innocence. As mentioned before, innocent boolean strategies are not closed under composition. One usually either imposes a further determinism constraint, or relaxes the innocence constraint. It might be instructive to see how trying to derive the boolean case from the set-based one using our methods directly points to the problem.

Indeed, suppose given any innocent game setting $(\mathbb{A}, \mathbb{P}, \mathbb{V}, i)$. We would like to show that two boolean polynomial functors, say

$$\overline{P_1}, \overline{P_2} \colon \widetilde{\mathbb{V}_{A,B} + \mathbb{V}_{B,C}} \to \widetilde{\mathbb{P}_{A,C}}$$

coincide. Here $\overline{P_1}$ is innocentisation followed by composition and $\overline{P_2}$ is the same, followed by another pass of innocentisation (and $\mathbb{P}$ could very well be replaced by $\mathbb{P}^+$). If we could show that each $\overline{P_i}$ factors as

$$\begin{array}{ccc} \widetilde{\mathbb{V}_{A,B} + \mathbb{V}_{B,C}} & \xrightarrow{P_i} & \widetilde{\mathbb{P}_{A,C}} \\ r_! \uparrow & & \downarrow l_! \\ \widehat{\mathbb{V}_{A,B} + \mathbb{V}_{B,C}} & \xrightarrow{P_i} & \widehat{\mathbb{P}_{A,C}}, \end{array}$$

where $r_!$ and $l_!$ are as in Proposition 2.16, then because we have already shown that $P_1 \cong P_2$, we would automatically get $\overline{P_1} \cong \overline{P_2}$ as desired.

Now, in all cases, both functors have the form

$$\widehat{\mathbb{V}_{A,B} + \mathbb{V}_{B,C}} \xrightarrow{Q} \widehat{\mathbb{P}_{A,B,C}} \xrightarrow{\Sigma_{\delta_1}} \widehat{\mathbb{P}_{A,C}} \xrightarrow{R} \widehat{\mathbb{P}_{A,C}},$$

where $\overline{Q}$ and $\overline{R}$ are only composed of $\overline{\Delta}$s and $\overline{\Pi}$s. Using the fact that $\Pi$s and $\Delta$s commute with $r_!$, we may thus hope to be able to prove the desired factorisation like so:

$$\begin{array}{ccccccc} \widehat{\mathbb{V}_{A,B} + \mathbb{V}_{B,C}} & \xrightarrow{Q} & \widehat{\mathbb{P}_{A,B,C}} & \xrightarrow{\Sigma_{\delta_1}} & \widehat{\mathbb{P}_{A,C}} & \xrightarrow{R} & \widehat{\mathbb{P}_{A,C}} \\ r_! \uparrow & & r_! \uparrow & ? & r_! \uparrow & & r_! \uparrow \searrow l_! \\ \widetilde{\mathbb{V}_{A,B} + \mathbb{V}_{B,C}} & \xrightarrow{\overline{Q}} & \widetilde{\mathbb{P}_{A,B,C}} & \xrightarrow{\overline{\Sigma_{\delta_1}}} & \widetilde{\mathbb{P}_{A,C}} & \xrightarrow{\overline{R}} & \widetilde{\mathbb{P}_{A,C}} = \widetilde{\mathbb{P}_{A,C}} \end{array}$$

where the right triangle commutes up to isomorphism because $l_! \dashv r_!$ is a reflection. The only problematic square is the marked one. And indeed, if the considered boolean strategies, say $X_1$ and $X_2$, are non-deterministic, $\Sigma_{\delta_1}(r_!(\overline{Q}[X_1, X_2]))$ may accept some $p \in \mathbb{P}_{A,C}$ in more than one way (see Harmer [15, 3.7.2]), which readily makes it non-isomorphic to any presheaf in the image of $r_!$.

**Remark 4.** *Exactly the same argument explains why boolean composition cannot agree with set-based composition in general (as was noted in Remark 2).*

Standardly, the problem is overcome by restricting to deterministic strategies, for which the marked square commutes. Again, this works indifferently in the ps or cs settings.

Finally, we proceed similarly in the case of copycat strategies. In this case, the problematic square

$$\begin{array}{ccc} \widehat{\mathbb{P}_A} & \xrightarrow{\Sigma_{\iota_0}} & \widehat{\mathbb{P}_{A,A}} \\ r_! \uparrow & & \uparrow r_! \\ \widetilde{\mathbb{P}_A} & \xrightarrow{\overline{\Sigma_{\iota_0}}} & \widetilde{\mathbb{P}_{A,A}} \end{array}$$

does commute, because the involved colimits are coproducts which are either empty or singleton:

**Proposition 4.17.** *In any innocent game setting* $(\mathbb{A}, \mathbb{P}, \mathbb{V}, i)$, *boolean copycat strategies are innocent. This extends to the setting of Proposition 4.16, so that, as is standard, copycats are innocent* pb-*strategies.*

*Proof.* By discrete fibredness of $\iota_0$, for all $p \in \mathbb{P}_{A,A}$, the comma category $p/\iota_0$ is either empty or connected. But $\sum_{\iota_0}(r_!(X))(p)$ is the colimit of

$$(p/\iota_0)^{op} \xrightarrow{\mathrm{cod}^{op}} \mathbb{P}_A^{op} \xrightarrow{X} 2 \xrightarrow{r} \mathrm{Set},$$

which is thus $\emptyset$ or $1$, hence isomorphic to $r_!(\overline{\sum_{\iota_0}(X)})(p)$. □

## 5 Conclusion and perspectives

We have introduced game settings and their innocent variant, a categorical framework for game semantics, with the hope of facilitating the construction of new game models. A lot remains to be done, starting with the incorporation of further instances. The saturated view of AJM games (Section 3.4) seems at hand, but will involve significantly more advanced category theory, as Street fibrations and stacks will replace discrete fibrations and sheaves. Less obvious is the treatment of more exotic game models [7, 11, 16, 28–31, 34], notably those based on event structures. We also plan to go beyond mere categories of games and strategies and construct structured categories of various kinds, depending on the considered language. These could be, e.g., cartesian closed, symmetric monoidal closed, or Freyd categories. Another direction is categorification: instead of reasoning up to isomorphism, we could refine our point of view and prove that games and strategies in fact form a bicategory, as, e.g., in Rideau and Winskel [34]. Finally, beyond game models, we should investigate game semantics, i.e., the correspondence with operational semantics, as initiated in Eberhart and Hirschowitz [12] in a different setting.

## A Categorical background

### A.1 Presheaves, restriction, extensions

This section is a brief introduction to the description of both adjoints to restriction, $\sum_F$ and $\prod_F$ (see Notation 1), in terms of coends and ends, respectively.

Consider the left adjoint first, $\sum_F$. It is determined up to canonical isomorphism by the *coend*

$$\sum_F(X)(d) \cong \int^{c \in \mathcal{C}} X(c) \times \mathcal{D}(d, F(c)),$$

which we abbreviate as

$$\int^{c \in \mathcal{C}} X(c) \times [d, F(c)], \qquad (4)$$

using the standard bracket notation for hom-sets (the ambient category should be clear from context). It will be enough to know that it is the quotient of the corresponding coproduct $\coprod_{c \in \mathcal{C}} X(c) \times$

$\mathcal{D}(d, F(c))$ by the smallest equivalence relation $\sim$ satisfying

$$(x, F(f) \circ g) \sim (x \cdot f, g)$$

for all $x \in X(c')$, $g \in \mathcal{D}(d, F(c))$ and $f \in \mathcal{C}(c, c')$. Graphically, this relation may be visualised by the following commuting 'diagram':

$$\begin{array}{c} \text{(diagram)} \end{array} \qquad (5)$$

We see that $f$ acts in different directions, on $X(c')$ and $[d, F(c)]$, and the equivalence relation takes this into account.

The right adjoint works dually: it is determined up to canonical isomorphism by the *end*

$$\prod\nolimits_F(X)(d) \cong \int_{c \in \mathcal{C}} \mathrm{Set}(\mathcal{D}(F(c), d), X(c)),$$

which we abbreviate as

$$\int_{c \in \mathcal{C}} [[F(c), d], X(c)]. \qquad (6)$$

In this case, the end admits a very simple description: the assignment $c \mapsto [F(c), d]$ defines a presheaf on $\mathcal{C}$, and the end is nothing but the set of natural transformations from this presheaf to $X$, so that we have

$$\prod\nolimits_F(X)(d) \cong \widehat{\mathcal{C}}([F(-), d], X)$$

(which is in fact an instance of the universal property for ends [22]).

Still, it may be instructive to realise that such natural transformations are equally the elements $\alpha$ of the product

$$\prod\nolimits_{c \in \mathcal{C}} [[F(c), d], X(c)]$$

satisfying a dual condition to (5), namely commutativity of

$$\begin{array}{c} \text{(diagram)} \end{array} \qquad (7)$$

for all $f : c \to c'$ and $g : F(c') \to d$, which is equivalent to naturality, i.e., commutativity of the square

$$\begin{array}{ccc} [F(c'), d] & \xrightarrow{\alpha_{c'}} & X(c') \\ {\scriptstyle [F(f), d]}\downarrow & & \downarrow{\scriptstyle X(f)} \\ [F(c), d] & \xrightarrow{\alpha_c} & X(c). \end{array}$$

Beware: unlike in (5), where an element of the coend is a pair of arrows (e.g., $(x \cdot f, g)$), an element of the end is here a natural family of arrows $1 \to X(c)$ indexed by all $g : F(c) \to d$.

### A.2 Fibrations

This section is a minimal introduction to (Grothendieck) fibrations and their relation to discrete fibrations.

**Definition A.1.** Consider any functor $p \colon \mathcal{E} \to \mathcal{C}$. A morphism $u \colon x \to y$ in $\mathcal{E}$ is *cartesian* when for all $v \colon z \to y$ in $\mathcal{E}$ and $k \colon p(z) \to p(x)$ making the bottom triangle commute

$$\begin{array}{c} \text{(diagram)} \end{array}$$

there exists a unique $w \colon z \to x$ such that $p(w) = k$ and $u \circ w = v$.

The functor $p$ is a *fibration* when for all objects $e \in \mathcal{E}$ and morphisms $f \colon c \to p(e)$ there exists a cartesian morphism $u \colon e' \to e$ such that $p(u) = f$. Such a morphism is called a *cartesian lifting* of $e$ along $f$.

Clearly, the discrete fibrations of Definition 2.8 are identical to fibrations whose cartesian liftings are unique, or equivalently whose *fibres* are discrete. The *fibre* of $p$ at any object $c \in \mathcal{C}$ is the subcategory of $\mathcal{E}$ spanning objects in $p^{-1}(c)$ and morphisms in $p^{-1}(id_c)$.

### A.3 Exact squares

An essential tool in our proofs will be Guitart's theory of *exact squares* [14], which we now recall.

**Definition A.2.** A *square* is a natural transformation

$$\begin{array}{ccc} A & \xrightarrow{T} & B \\ S\downarrow & \stackrel{\varphi}{\Rightarrow} & \downarrow V \\ C & \xrightarrow{U} & D. \end{array} \qquad (8)$$

of small categories and functors.

Any square yields by restriction a square as in the middle below, and so by adjunction (the so-called mate calculus) two further squares as on the left and right:

$$\begin{array}{c} \text{(three diagrams)} \end{array}$$

**Definition A.3.** A square $\varphi$ is *exact* if and only if $\sum_\varphi$ is an isomorphism.

The following is well-known:

**Property 1.** *A square $\varphi$ is exact if and only if $\prod_\varphi$ is an isomorphism.*

The notion of exactness thus corresponds to commutation up to isomorphism of the two diagrams on the sides above.

**Remark 5.** *In an attempt to spare the reader a few headaches trying to remember the directions of $\sum$'s, $\Delta$'s and $\prod$'s, as well as of the induced natural transformations, let us share our mnemonic: the original transformation $\varphi$ points to the $\sum$'ed functor if we are to reason about $\sum_\varphi$, and from the $\prod$'ed functor if we are to reason about $\prod_\varphi$. Furthermore, induced natural transformations flow along left adjoints and against right adjoints (hence along $\sum$s for $\sum_\varphi$, and against $\prod$s for $\prod_\varphi$).*

**Notation 3.** *In the proofs, we will manipulate diagrams of restriction and its adjoints. In order to reduce notational clutter, hats will be omitted and arrows will point in the direction of underlying functors. This means that $\Delta$ arrows will point in the "wrong" direction, in the sense that if $F\colon X \to Y$, we will write $Y \xleftarrow{\Delta_F} X$ for $\widehat{Y} \xrightarrow{\Delta_F} \widehat{X}$.*

Let us conclude this section with a few basic lemmas about exact squares. First, we recall that a square (8) is a *comma square* when it is the terminal lax cone to $(U, V)$. More formally, it is a terminal object in the category whose objects are spans $C \xleftarrow{S'} A' \xrightarrow{T'} B$ equipped with a natural transformation $\varphi'\colon US' \to VT'$, and whose morphisms $(A', S', T', \varphi') \to (A'', S'', T'', \varphi'')$ are functors $m\colon A' \to A''$ such that $S'' \circ m = S'$, $T'' \circ m = T'$, and $\varphi'' \circ m = \varphi'$ (where $\circ$ here denotes whiskering). *Cocomma squares* are defined exactly dually, i.e., with $D$, $U$ and $V$ varying instead of $A$, $S$, and $T$. Guitart shows:

**Lemma A.4.** *Any comma (resp. cocomma) square is exact.*

*Proof.* See Guitart [14], Examples 1.14.2 and 1.14.3. □

**Lemma A.5.** *For any functor $f\colon A \to B$, the square below left is exact; furthermore, the square below right is exact iff $f$ is fully faithful:*

$$\begin{array}{ccc} A \xrightarrow{f} B & & A =\!=\!= A \\ \| \quad \overset{id_f}{\Rightarrow} \quad \| & & \| \quad \overset{id_f}{\Rightarrow} \quad \downarrow f \\ A \xrightarrow{f} B & & A \xrightarrow{f} B \end{array}$$

*Proof.* See Guitart [14], Examples 1.14.1 and 1.14.4. □

**Lemma A.6.** *Exact squares are stable under horizontal composition.*

*Proof.* See Guitart [14], Theorem 1.8. □

**Lemma A.7.** *Any square (8) in which $\varphi$ and $S$ are identities and $V$ is fully faithful is exact.*

*Proof.* We obtain the given square as the horizontal composite

$$\begin{array}{c} A \xrightarrow{T} B =\!=\!= B \\ \| \quad T \quad \| \quad \quad \downarrow V \\ A \xrightarrow{T} B \xrightarrow{V} C, \\ \underset{U}{\longrightarrow} \end{array}$$

which is exact by Lemma A.6, because both squares are exact by Lemma A.5. □

Guitart provides the following necessary and sufficient criterion for exactness:

**Lemma A.8** (Guitart [14], Theorem 1.2 (zig-zag criterion)). *A square as in (8) is exact if and only if, for all objects $c$ of $C$, $d$ of $D$, and morphisms $f\colon Uc \to Vd$:*

- *there is an object $a$ of $A$ such that $f$ factors as $Uc \xrightarrow{Ug} USa \xrightarrow{\varphi_a} VTa \xrightarrow{Vh} Vd$,*

- *and for all two such factorisations $Uc \xrightarrow{Ug} USa \xrightarrow{\varphi_a} VTa \xrightarrow{Vh} Vd$ and $Uc \xrightarrow{Ug'} USa' \xrightarrow{\varphi_{a'}} VTa' \xrightarrow{Vh'} Vd$, there exists a commuting diagram (which Guitart calls a lantern)*

$$\begin{array}{c} c \\ Sa \xleftarrow{g} \cdot \xleftarrow{St_0} \cdot \xrightarrow{St_1} \cdots \xleftarrow{St_{n-1}} \cdot \xrightarrow{St_n} Sa' \\ Ta \xrightarrow{Tt_0} \cdot \xrightarrow{Tt_1} \cdots \xrightarrow{Tt_{n-1}} \cdot \xrightarrow{Tt_n} Ta' \\ h \searrow \quad d. \quad \swarrow h' \end{array}$$

Let us conclude this section with an apparently new (though not surprising) result showing that pullbacks of (op)fibrations are exact squares (not in the same direction!).

**Lemma A.9.** *Any pullback square (8) with $V$ a fibration is exact. Similarly, if $U$ is an opfibration, then (8) is exact.*

*Proof.* We proceed by the zig-zag criterion. Consider any $f\colon Uc \to Vb$. Let us first establish existence of the desired factorisation, by considering any cartesian lifting $l\colon b_0 \to b$ of $b$ along $f$. We obtain $Uc = Vb_0$, hence by universal property of pullback (By the standard categorification/nerve adjunction, limits of categories are computed as limits of underlying simplicial sets, hence pointwise.) a unique $a$ such that $c = Sa$ and $b_0 = Ta$. The original $f$ thus factors as

$$Uc \xrightarrow{U(id)} USa \xrightarrow{id} VTa \xrightarrow{Vl} Vb.$$

We now need to show that any factorisation is connected to this one by some lantern. So consider any factorisation of $f$ as

$$Uc \xrightarrow{Ug} USa' \xrightarrow{id} VTa' \xrightarrow{Vh} Vb.$$

Because fibrations are stable under pullback, $S$ is a fibration, so we may pick a cartesian lifting, say $k\colon a'' \to a'$, of $a'$ along $g$, so that in particular $Sa'' = c$ and $Sk = g$. Now, we have

$$VTa'' = USa'' = Uc = USa = VTa$$

hence a commuting triangle as below left

$$\begin{array}{cc} VTa'' \xrightarrow{VTk} VTa' \xrightarrow{Vh} Vb & \quad Ta'' \xrightarrow{Tk} Ta' \xrightarrow{h} b \\ \| \quad \quad \quad \nearrow Vl & \quad u\downarrow \quad \quad \nearrow l \\ VTa & \quad Ta \end{array}$$

(because $VTk = USk = Ug$ so $Vh \circ VTk = f = Vl$). So by cartesianness of $l$, we obtain a (unique) $u$ making the above right triangle commute. We thus obtain the following commuting diagrams:

$$\begin{array}{c} c \\ Sa =\!=\!= Sa'' \xrightarrow{Sk} Sa' \\ \quad \quad \quad g \nearrow \end{array}$$

$$\begin{array}{c} Ta \xleftarrow{u} Ta'' \xrightarrow{Tk} Ta' \\ l \searrow \downarrow \swarrow h \\ b. \end{array}$$

But since $Vu = id$, we have that $U(id) = Vu$ so by universal property of pullback again there exists a unique $w\colon a'' \to a$ such that $Sw = id$ and $Tw = u$. The above diagram thus yields a lantern, as desired. □

### A.4 Distributivity

As we saw, exact squares ensure commutation of restriction with left (resp. right) extension. In some of our proofs, we will also encounter squares for which we will need left extension to commute with right extension. This is an instance of what we will call distributive squares, which we now introduce.

Consider any exact square as (8). The inverse of $\sum_\varphi$ yields a natural transformation as below left

$$\begin{array}{ccc} \widehat{A} & \xleftarrow{\Delta_T} & \widehat{B} \\ \Sigma_S \downarrow & \overset{\sum_\varphi^{-1}}{\Leftarrow} & \downarrow \Sigma_V \\ \widehat{C} & \xleftarrow{\Delta_U} & \widehat{D} \end{array} \qquad \begin{array}{ccc} \widehat{A} & \xrightarrow{\Pi_T} & \widehat{C} \\ \Sigma_S \downarrow & \overset{\tilde{\varphi}}{\Rightarrow} & \downarrow \Sigma_V \\ \widehat{B} & \xrightarrow{\Pi_U} & \widehat{D}, \end{array}$$

which by adjunction induces a natural transformation as on the right.

**Definition A.10.** An exact square is *distributive* when $\tilde{\varphi}$ is an isomorphism.

Here is a useful construction of distributive squares. Recalling Proposition 4.6, consider any functors $A \xrightarrow{S} B \xrightarrow{U} D$ with $S$ a discrete fibration $U$ fully faithful. Let $C = \text{el}(\prod_U(\text{sing}(S)))$ denote the category of elements of the right extension of $\text{sing}(S)$ along $U$, and let $V$ denote its projection to $D$. Because extension along a fully faithful functor is fully faithful, the counit of the adjunction $\Delta_U \dashv \prod_U$ is an isomorphism. This yields a pullback square

$$\begin{array}{ccc} A & \xrightarrow{T_{S,U}} & D/\prod_U(\text{sing}(S)) \\ S \downarrow & \lrcorner & \downarrow p_{\prod_U(\text{sing}(S))} \\ B & \xrightarrow{U} & D \end{array} \qquad (9)$$

which is exact by Lemma A.9.

**Definition A.11.** Any square obtained in this way is called a *local pushforward* square.

**Lemma A.12.** *Local pushforward squares are distributive.*

*Proof.* Consider any local pushforward square (8). We prove the corresponding result for the equivalent categories of discrete fibrations. There, because $V$ and hence $S$ are discrete fibrations, $\Sigma_S$ becomes post-composition with $S$, and similarly for $\Sigma_V$. So the result essentially states that for any discrete fibration $P: E \to A$,

$$\prod_U(SP) \cong \prod_U(S) \circ \prod_T(P) \cong V \circ \prod_T(P),$$

where the first isomorphism follows from a simple diagram chase showing that $\prod_U(S) \circ \prod_T(P)$ enjoys the universal property of $\prod_U(SP)$. □

## B Main proofs

### B.1 Adequacy

*Proof of Proposition 2.10.* First, because $\prod_!$ is right adjoint to $\Delta_!$, it preserves the terminal object (which is the unique object of $\widehat{\emptyset}$), hence maps 1 to the terminal presheaf on $\mathbb{P}_A$, defined to map any play in $\mathbb{P}_A$ to 1. So we reduce to showing that $id_A \cong \sum_{\iota_0}(1)$.

Discrete fibredness of $\iota_0$ entails that left extension may be expressed as a mere coproduct: we have, for any presheaf $X$ on $\mathbb{P}_A$ and $p \in \mathbb{P}_{A,A}$,

$$(\sum_{\iota_0} X)(p) \cong \sum_{\{q \in \mathbb{P}_A | \iota_0(q) = p\}} X(q).$$

So in particular when $X = 1$ we get

$$\sum_{\iota_0}(1)(p) \cong \sum_{\{q \in \mathbb{P}_A | \iota_0(q) = p\}} 1 \cong \begin{cases} 1 & \text{if } p \in \text{Im}(\iota_0) \\ \emptyset & \text{otherwise,} \end{cases}$$

as desired. □

*Proof of Proposition 2.13.* Starting from the end, consider $\sum_{\delta_1}$. Thanks to discrete fibredness of $\delta_1$, we have for any presheaf $X''$ on $\mathbb{P}_{A,B,C}$ and $p \in \mathbb{P}_{A,C}$:

$$\sum_{\delta_1}(X'')(p) \cong \sum_{\{u \in \mathbb{P}_{A,B,C} | \delta_1(u) = p\}} X''(u).$$

The next step is to unfold the right extension $\prod_{[id, id]}$. In this case, $[id, id]$ is easily seen to be a discrete opfibration, which entails that right extension may be expressed as the product

$$\prod_{[id, id]}(X')(u) \cong \prod_{\{v \in (\mathbb{P}_{A,B,C} + \mathbb{P}_{A,B,C}) | [id, id](v) = u\}} X'(v),$$

i.e., unfolding $X'$ as the copairing $[X'_0, X'_1]$, $v$ as the pair $(i, u')$ and using the constraint $[id, id](v) = [id, id](i, u') = u' = u$:

$$\prod_{[id, id]}[X'_0, X'_1](u) \cong \prod_{i \in \{0,1\}}(X'_i(u)) \cong X'_0(u) \times X'_1(u).$$

Finally, because $\Delta_{\delta_2 + \delta_0}[X_0, X_1] = [\Delta_{\delta_2}(X_0), \Delta_{\delta_0}(X_1)]$, we obtain that the candidate composition of Definition 2.12 maps any $[X_0, X_1]$ to

$$\begin{aligned} &\sum_{\delta_1}(\prod_{[id, id]}(\Delta_{\delta_2 + \delta_0}[X_0, X_1]))(p) \\ &= \sum_{\{u \in \mathbb{P}_{A,B,C} | \delta_1(u) = p\}} \prod_{[id, id]}[\Delta_{\delta_2}(X_0), \Delta_{\delta_0}(X_1)](u) \\ &= \sum_{\{u \in \mathbb{P}_{A,B,C} | \delta_1(u) = p\}} \Delta_{\delta_2}(X_0)(u) \times \Delta_{\delta_0}(X_1)(u) \\ &= \sum_{\{u \in \mathbb{P}_{A,B,C} | \delta_1(u) = p\}} X_0(\delta_2(u)) \times X_1(\delta_0(u)), \end{aligned}$$

which clearly coincides with Tsukada and Ong's definition. □

### B.2 Associativity

In this section, we prove Theorem 2.15. Let us start with an alternative description of composition, which relies on the following intermediate category:

**Definition B.1.** For any triple of arenas $A$, $B$, $C$, let $\mathbb{P}_{(A,B),(B,C)}$ denote the *lax colimit* [22] of

$$\mathbb{P}_{A,B} \longleftarrow \mathbb{P}_{A,B,C} \longrightarrow \mathbb{P}_{B,C},$$

i.e., the initial category equipped with natural transformations

$$\begin{array}{ccc} \mathbb{P}_{A,B} & \xleftarrow{\phantom{xxx}} \mathbb{P}_{A,B,C} \xrightarrow{\phantom{xxx}} & \mathbb{P}_{B,C} \\ & \searrow \overset{\lambda}{\Rightarrow} \downarrow \overset{\rho}{\Leftarrow} \swarrow & \\ & \mathbb{P}_{(A,B),(B,C)}. & \end{array}$$

The obtained category has as objects the disjoint union of objects from $\mathbb{P}_{A,B}$, $\mathbb{P}_{B,C}$, and $\mathbb{P}_{A,B,C}$. It inherits the corresponding morphisms, and has additional morphisms $\lambda_u: \delta_2 u \to u$ and $\rho_u: \delta_0 u \to u$ for all $u$ satisfying the obvious naturality requirements:

$$\lambda_{u'} \circ \delta_2 f = f \circ \lambda_u \qquad \text{and} \qquad \rho_{u'} \circ \delta_0 f = f \circ \rho_u,$$

for all $f: u \to u'$ in $\mathbb{P}_{A,B,C}$. We have:

**Proposition B.2.** *Composition is isomorphic to the polynomial functor*

$$\mathbb{P}_{A,B} + \mathbb{P}_{B,C} \xrightarrow{\Pi} \mathbb{P}_{(A,B),(B,C)} \xleftarrow{\Delta} \mathbb{P}_{A,B,C} \xrightarrow{\Sigma} \mathbb{P}_{A,C}.$$

*Proof.* This will follow from Lemma A.4 by observing that the square

$$
\begin{array}{ccc}
\mathbb{P}_{A,B,C} + \mathbb{P}_{A,B,C} & \xrightarrow{\nabla} & \mathbb{P}_{A,B,C} \\
{\scriptstyle \delta_2 + \delta_0} \downarrow & \Longrightarrow & \downarrow \\
\mathbb{P}_{A,B} + \mathbb{P}_{B,C} & \longrightarrow & \mathbb{P}_{(A,B),(B,C)}
\end{array}
$$

is a cocomma square, where $\nabla$ denotes the copairing $[id, id]$. Indeed, both categories have the same universal property, expressed differently. □

Proving associativity thus reduces to showing that the perimeter of the top diagram of Figure 1 commutes up to isomorphism. In order to do this, we introduce the category $\mathbb{P}_{(A,B),(B,C),(C,D)}$, similar to $\mathbb{P}_{(A,B),(B,C)}$ but with four arenas, which is constructed as the lax colimit of

$$
\mathbb{P}_{A,B} \leftarrow \mathbb{P}_{A,B,C} \xrightarrow{\mathbb{P}_{B,C}} \mathbb{P}_{B,C,D} \rightarrow \mathbb{P}_{C,D}
$$

$$
\mathbb{P}_{A,B,C,D}.
$$

In the top diagram of Figure 1, both little squares commute up to isomorphism because the underlying squares do. It thus suffices to show that both heptagons commute up to isomorphism. Both cases are symmetric, so we only treat the bottom left one. We then need to introduce yet another category, $\mathbb{P}_{(A,B),(B,C,D)}$, which is like $\mathbb{P}_{(A,B),(B,C),(C,D)}$, except that $\mathbb{P}_{B,C,D}$ is not decomposed into $\mathbb{P}_{B,C}$ and $\mathbb{P}_{C,D}$: it is the lax colimit of

$$
\mathbb{P}_{A,B} \leftarrow \mathbb{P}_{A,B,C,D} \rightarrow \mathbb{P}_{B,C,D}.
$$

These categories are related by full embeddings

$$
\mathbb{P}_{(A,B),(B,C)} \hookrightarrow \mathbb{P}_{(A,B),(B,C),(C,D)} \hookleftarrow \mathbb{P}_{(A,B),(B,C,D)}.
$$

The crucial reason why the heptagon commutes is:

**Lemma B.3.** *The square*

$$
\begin{array}{ccc}
\mathbb{P}_{A,B} + \mathbb{P}_{B,C,D} & \longrightarrow & \mathbb{P}_{(A,B),(B,C,D)} \\
{\scriptstyle \mathbb{P}_{A,B} + \delta_1} \downarrow & & \downarrow {\scriptstyle \mathbb{P}_{(A,B),\delta_1}} \\
\mathbb{P}_{A,B} + \mathbb{P}_{B,D} & \xrightarrow{[\lambda,\rho]} & \mathbb{P}_{(A,B),(B,D)}
\end{array}
$$

*is a local pushforward square, i.e., it is a pullback, $\mathbb{P}_{A,B} + \delta_1$ is a discrete fibration, $[\lambda, \rho]$ is fully faithful, and*

$$
\mathbb{P}_{(A,B),\delta_1} \cong \prod_{[\lambda,\rho]} (\mathbb{P}_{A,B} + \delta_1).
$$

*Proof sketch.* Everything is direct, except the last isomorphism, which says that $\mathbb{P}_{(A,B),(B,C,D)}$ is a sheaf for the topology induced by the full embedding $\mathbb{P}_{A,B} + \mathbb{P}_{B,D} \hookrightarrow \mathbb{P}_{(A,B),(B,D)}$. This reduces to the fact that any $u \in \mathbb{P}_{A,B,C,D}$ is entirely determined by giving its projections $u_1 = \delta_2(\delta_2(u)) \in \mathbb{P}_{A,B}$ and $u_2 = \delta_0(\delta_2(u)) \in \mathbb{P}_{B,D}$, plus some $p \in \mathbb{P}_{B,C,D}$ such that $\delta_1(p) = u_2$, which holds by zipping.

In more detail, the considered topology has canonical sieves: for objects in the image of the embedding, the canonical sieve is just the total sieve; for $u \in \mathbb{P}_{A,B,D}$, the canonical sieve is generated by the cospan

$$
\delta_2(u) \rightarrow u \leftarrow \delta_0(u).
$$

Thus, the sheaf condition for $\mathbb{P}_{(A,B),(B,C,D)}$ reduces to saying that, implicitly coercing discrete fibrations $p: \mathbb{E} \rightarrow \mathbb{P}_{(A,B),(B,C)}$ into the corresponding presheaf $\mathrm{sing}((\mathbb{)E}, p)$, the restriction map

$$
\mathbb{P}_{(A,B),(B,C,D)}(u) \rightarrow \mathbb{P}_{A,B}(\delta_2(u)) \times \mathbb{P}_{B,C,D}(\delta_0(u))
$$

is an isomorphism. But $\mathbb{P}_{A,B}(\delta_2(u)) \cong 1$, so this further reduces to the restriction map

$$
\mathbb{P}_{(A,B),(B,C,D)}(u) \rightarrow \mathbb{P}_{B,C,D}(\delta_0(u))
$$

being an isomorphism. This is only non-trivial when $u \in \mathbb{P}_{A,B,D}$ and in that case it is equivalent to
$\forall u \in \mathbb{P}_{A,B,D}, v \in \mathbb{P}_{B,C,D}, (\delta_0(u) = \delta_1(v))$
$\Rightarrow \exists! w \in \mathbb{P}_{A,B,C,D}, \delta_2(w) = u \wedge \delta_0(w) = v,$
i.e., the left square of (3) being a pullback on objects. □

This leads us to fill the heptagon as at the bottom of Figure 1. The top right square commutes by Lemmas A.12 and B.3. The top triangle commutes up to isomorphism by Lemma A.7, the bottom one because the underlying diagram commutes, and the bottom-right square by Lemma A.9.

Associativity finally follows from:

**Lemma B.4.** *The left square commutes up to isomorphism.*

*Proof.* By the classical limit formula for right extensions, given any presheaf $X$ on $\mathbb{P}_{A,B} + \mathbb{P}_{(B,C),(C,D)}$,

- following the left and bottom arrows we obtain a presheaf $X'$ mapping any $u \in \mathbb{P}_{A,B,C,D}$ to the limit of

$$
\begin{array}{c}
((\mathbb{P}_{A,B} + \mathbb{P}_{(B,C),(C,D)})/u)^{op} \\
\downarrow \\
(\mathbb{P}_{A,B} + \mathbb{P}_{(B,C),(C,D)})^{op} \\
\downarrow X \\
\mathrm{Set};
\end{array}
$$

- following the top and right arrows, we obtain a presheaf $X'$ mapping any $u$ to a similar limit with $\mathbb{P}_{A,B} + \mathbb{P}_{(B,C),(C,D)}$ replaced by $\mathbb{P}_{A,B} + \mathbb{P}_{B,C,D}$.

Now, the inclusion functor

$$
(\mathbb{P}_{A,B} + \mathbb{P}_{B,C,D})/u \hookrightarrow (\mathbb{P}_{A,B} + \mathbb{P}_{(B,C),(C,D)})/u
$$

is readily checked to be final [25, IX.3], so its opposite is initial and both limits are isomorphic. □

### B.3 Unitality

Left and right unitality are entirely symmetric, so we only treat one. First, we observe that, because $\iota_0: \mathbb{P}_A \rightarrow \mathbb{P}_{A,A}$ is a discrete fibration, so is $\iota_0 + \mathbb{P}_{A,B}: \mathbb{P}_A + \mathbb{P}_{A,B} \rightarrow \mathbb{P}_{A,A} + \mathbb{P}_{A,B}$. Consider the diagram:

$$
\begin{array}{ccccc}
\emptyset + \mathbb{P}_{A,B} & \xrightarrow{\Pi} & \mathbb{P}_A + \mathbb{P}_{A,B} & \xrightarrow{\Sigma_{\iota_0 + \mathbb{P}_{A,B}}} & \mathbb{P}_{A,A} + \mathbb{P}_{A,B} \\
& \searrow{\scriptstyle \Pi} & \downarrow{\scriptstyle \Pi} & & \downarrow{\scriptstyle \Pi_{[\lambda,\rho]}} \\
& & \mathbb{C} & \xrightarrow{\Sigma} & \mathbb{P}_{(A,A),(A,B)} \\
& & {\scriptstyle \Delta} \uparrow & & \uparrow{\scriptstyle \Delta} \\
& & \mathbb{P}_{A,B} & \xrightarrow{\Sigma} & \mathbb{P}_{A,A,B} & \xrightarrow{\Sigma} & \mathbb{P}_{A,B}
\end{array}
$$

where the functor $\mathbb{C} \rightarrow \mathbb{P}_{(A,A),(A,B)}$ is defined as

$$
\mathrm{el}(\prod_{[\lambda,\rho]} (\mathrm{sing}(\iota_0 + \mathbb{P}_{A,B}))).
$$

**Figure 1.** Diagram for associativity and zoom into the bottom left heptagon

We thus obtain a local pushforward square which commutes up to isomorphism by Lemma A.12.

Let us now show that the bottom square is a pullback. By construction, an object $c$ of $\mathbb{C}$ over any $u \in \mathbb{P}_{A,A,B}$ consists of an object $c_1 \in \mathbb{P}_A$ over $\delta_2(u)$ and an object $c_2 \in \mathbb{P}_{A,B}$ over $\delta_0(u)$. The latter has to be $\delta_0(u)$, and the former has to be $\delta_0(\delta_2(u))$ or equivalently $\delta_1(\delta_0(u))$, but it exists only when $u \cong \iota_0(\delta_0(u))$, i.e., when $u$ is in the image of $\mathbb{P}_{A,B} \to \mathbb{P}_{A,A,B}$. Thus, the bottom square is exact by Lemma A.9.

Finally, the bottom row is isomorphic the identity on $\mathbb{P}_{A,B}$, so it only remains to prove that the composite $\mathbb{P}_{A,B} \cong \emptyset + \mathbb{P}_{A,B} \xrightarrow{\Pi} \mathbb{C} \xleftarrow{\Delta} \mathbb{P}_{A,B}$ is also isomorphic to the identity, which follows from the standard end formula for $\prod$ and the explicit description of lax colimits.

### B.4 The boolean case

In this section, we show that our results in the presheaf case transfer to the boolean case. This will follow from showing that all the polynomial functors that we used commute with booleanisation. This is easy for left extensions and restrictions:

**Proposition B.5.** *For all functors $F: \mathbb{C} \to \mathbb{D}$, the following squares commute up to isomorphism.*

$$\begin{array}{ccc} [\mathbb{C}^{op}, \text{Set}] & \xrightarrow{\sum_F} & [\mathbb{D}^{op}, \text{Set}] \\ {!}_! \downarrow & & \downarrow {!}_! \\ [\mathbb{C}^{op}, 2] & \xrightarrow{\overline{\sum}_F} & [\mathbb{D}^{op}, 2] \end{array} \qquad \begin{array}{ccc} [\mathbb{C}^{op}, \text{Set}] & \xleftarrow{\Delta_F} & [\mathbb{D}^{op}, \text{Set}] \\ {!}_! \downarrow & & \downarrow {!}_! \\ [\mathbb{C}^{op}, 2] & \xleftarrow{\overline{\Delta}_F} & [\mathbb{D}^{op}, 2] \end{array}$$

*Proof.* Commutation with restriction is obvious. For left extension, the reflection $!$, being a left adjoint, preserves colimits, which is precisely what $\sum_F$ computes. □

Things do not work out so well with right extensions in general. In order to show that our polynomial functors commute with booleanisation, it is thus useful to delineate a sufficiently large class of limits that are preserved by $!$:

**Lemma B.6.** *The left adjoint $!$ preserves products and the terminal object.*

*Proof.* That $!$ preserves $1$ is obvious. Now, a product $\prod_i X_i$ is non-empty just when each $X_i$ is, hence just when $!(X_i) = 1$ for all $i$, i.e., when $\prod_i !(X_i) = 1$. Thus $!(\prod_i X_i) = 1$ iff $\prod_i !(X_i) = 1$, hence $!(\prod_i X_i) = \prod_i !(X_i)$. □

**Proposition B.7.** *Booleanisation commutes with $\prod_F$, for any $F: \mathbb{C} \to \mathbb{D}$ such that for all $d \in \text{ob}(\mathbb{D})$ the comma category $F/d$ is a coproduct of categories with a terminal object.*

*Proof.* Indeed, consider any such $F$. For any $X \in \widehat{\mathbb{C}}$ and $d \in \mathbb{D}$, letting $F/d \cong \sum_i \mathbb{D}_i^d$ with $\varphi_i^d: F(c_i^d) \to d$ denoting the terminal object in $\mathbb{D}_i^d$, we have

$$\begin{array}{rcl} !((\prod_F(X))(d)) & \cong & !(\lim((F/d)^{op} \to \mathbb{C}^{op} \xrightarrow{X} \text{Set})) \\ & \cong & !(\prod_i X(c_i^d)) \\ & \cong & \prod_i !(X(c_i^d)) \qquad \text{(by Lemma B.6)} \\ & \cong & (\lim((F/d)^{op} \to \mathbb{C}^{op} \xrightarrow{X} \text{Set} \xrightarrow{!} 2)) \\ & \cong & \prod_F(! \circ X)(d), \end{array}$$

as desired. □

**Proposition B.8.** *The class of functors $F$ such that $\prod_F$ commutes with booleanisation contains all functors $\nabla_{\mathbb{C}}: \mathbb{C} + \mathbb{C} \to \mathbb{C}$ and $!: \emptyset \to \mathbb{C}$, and it is stable under composition and coproduct (i.e., $F + G: \mathbb{C} + \mathbb{C}' \to \mathbb{D} + \mathbb{D}'$ is in it if $F$ and $G$ are).*

*Proof.* Easy consequences of the previous proposition. □

*Proof of Proposition 2.18.* Both results state that two polynomial functors, say $\overline{P_1}$ and $\overline{P_2}$ between categories of the form $[\mathbb{C}^{op}, 2]$ are naturally isomorphic. But knowing that their set-versions, say $P_1$ and $P_2$, are isomorphic, we may form

$$\begin{array}{ccc} [\mathbb{C}^{op}, \text{Set}] & \xrightarrow[P_2]{P_1} & [\mathbb{D}^{op}, \text{Set}] \\ {!}_! \downarrow & & \downarrow {!}_! \\ [\mathbb{C}^{op}, 2] & \xrightarrow[\overline{P_2}]{\overline{P_1}} & [\mathbb{D}^{op}, 2]. \end{array}$$

In both cases, this diagram commutes serially by Propositions B.5 and B.8, and we have proved that the top parallel functors are isomorphic. But $[\mathbb{C}^{op}, \mathbb{I}]$ is epi, which entails that the bottom parallel functors are also isomorphic, as desired. □

## B.5 Constraining strategies

The goal of this section is to prove Proposition 3.18. Before attacking this, let us reformulate the locality condition on c, using the category $\mathbb{P}_{(A,B),(B,C)}$ of Definition B.1.

**Lemma B.9.** *An embedding* $c\colon \mathbb{P}^c \hookrightarrow \mathbb{P}$ *as in Proposition 3.18, i.e., between game settings sharing the same set of arenas and whose components are discrete fibrations, is a local constraint iff* $\mathrm{sing}(c_{(A,B),(B,C)})$, *the presheaf induced by the embedding* $\mathbb{P}^c_{(A,B),(B,C)} \hookrightarrow \mathbb{P}_{(A,B),(B,C)}$, *is a sheaf for the Grothendieck topology induced by the embedding* $\mathbb{P}_{A,B} + \mathbb{P}_{B,C} \hookrightarrow \mathbb{P}_{(A,B),(B,C)}$.

*Proof.* A direct check. □

*Proof of Proposition 3.18.* By Proposition B.2, the result reduces to the commutation of

$$\begin{array}{ccccccc}
\mathbb{P}^c_{A,B} + \mathbb{P}^c_{B,C} & \xrightarrow{\Pi} & \mathbb{P}^c_{(A,B),(B,C)} & \xleftarrow{\Delta} & \mathbb{P}^c_{A,B,C} & \xrightarrow{\Sigma} & \mathbb{P}^c_{A,C} \\
{\scriptstyle \Sigma_c \times \Sigma_c}\downarrow & & \downarrow{\scriptstyle \Sigma_c} & & \downarrow{\scriptstyle \Sigma_c} & & \downarrow{\scriptstyle \Sigma_c} \\
\mathbb{P}_{A,B} + \mathbb{P}_{B,C} & \xrightarrow[\Pi]{} & \mathbb{P}_{(A,B),(B,C)} & \xleftarrow[\Delta]{} & \mathbb{P}_{A,B,C} & \xrightarrow[\Sigma]{} & \mathbb{P}_{A,C}
\end{array}$$

up to isomorphism. The right-hand square commutes up to isomorphism because the underlying square does; the middle one commutes by Lemma A.9; and the left-hand one by hypothesis and Lemma A.12. □

## B.6 Concurrent innocence

In this section, we prove Theorem 4.13, in two lemmas. The first states stability of innocence under composition; the second says that copycat strategies are innocent.

**Lemma B.10.** *In any innocent game setting, the composite of two innocent strategies is again innocent.*

*Proof.* Because a strategy $X$, say on $(A, B)$, is innocent iff it is isomorphic to $\prod_{i_{A,B}}(\Delta_{i_{A,B}}(X))$, it suffices to show that starting from any pair of behaviours $[B_1, B_2] \in \widehat{\mathbb{V}_{A,B} + \mathbb{V}_{B,C}}$, if we extend them to innocent strategies, compose, and eventually apply innocentisation (i.e., $\prod_{i_{A,C}} \circ \Delta_{i_{A,C}}$), then the last step is redundant. In other words, we need to show that the perimeter of

$$\begin{array}{ccccccc}
& & \mathbb{P}_{A,B} + \mathbb{P}_{B,C} & \xrightarrow{\Pi} & \mathbb{P}_{(A,B),(B,C)} & \xleftarrow{\Delta} & \mathbb{P}_{A,B,C} & \xrightarrow{\Sigma} & \mathbb{P}_{A,C} \\
& \nearrow{\scriptstyle \Pi} & & & \uparrow{\scriptstyle \Delta} & & \uparrow{\scriptstyle \Delta} & & \uparrow{\scriptstyle \Delta} \\
\mathbb{V}_{A,B} + \mathbb{V}_{B,C} & \xrightarrow{\Pi} & \mathbb{V}_{(A,B),(B,C)} & \xleftarrow{\Delta} & \mathbb{V}_{A,B,C} & \xrightarrow{\Sigma} & \mathbb{V}_{A,C} & & \quad (10) \\
& \searrow{\scriptstyle \Pi} & & & \downarrow{\scriptstyle \Pi} & & \downarrow{\scriptstyle \Pi} & & \downarrow{\scriptstyle \Pi} \\
& & \mathbb{P}_{A,B} + \mathbb{P}_{B,C} & \xrightarrow[\Pi]{} & \mathbb{P}_{(A,B),(B,C)} & \xleftarrow[\Delta]{} & \mathbb{P}_{A,B,C} & \xrightarrow[\Sigma]{} & \mathbb{P}_{A,C}
\end{array}$$

commutes up to isomorphism. We proceed by showing that innocentisation is redundant at every intermediate step, but this requires us to define intermediate categories of views adequately: $\mathbb{V}_{ABC}$ is defined as the pullback $\mathbb{P}_{A,B,C} \times_{\mathbb{P}_{A,C}} \mathbb{V}_{A,C}$, i.e., the full subcategory of $\mathbb{P}_{A,B,C}$ spanning plays whose $\delta_1$-projection is a view; for $\mathbb{V}_{(A,B),(B,C)}$, consider first the lax colimit $\mathbb{C}$ of $\mathbb{P}_{A,B} \leftarrow \mathbb{V}_{A,B,C} \to \mathbb{P}_{B,C}$; $\mathbb{V}_{(A,B),(B,C)}$ is its full subcategory spanning objects from $\mathbb{V}_{A,B,C}$, $\mathbb{V}_{A,B}$ and $\mathbb{V}_{B,C}$.

Returning to our claim, the top-left square commutes by Lemma A.7, the top square because the underlying diagram commutes, the top-right square by Lemma A.9, the bottom-left square because the underlying diagram commutes, the bottom-right one by locality. Finally, the bottom square commutes because it is exact by the zig-zag criterion using view-analyticity and observing that it is a pullback of fully faithful functors. □

We finally prove innocence of identities:

**Lemma B.11.** *Copycat strategies are innocent.*

*Proof.* We proceed as for preservation of innocence by composition: by showing that copycat is the same as copycat followed by innocentisation. This yields the diagram

$$\begin{array}{ccc}
& \mathbb{P}_A \xrightarrow{\Sigma} \mathbb{P}_{A,A} & \\
\nearrow{\scriptstyle \Pi} \uparrow{\scriptstyle \Delta} & & \uparrow{\scriptstyle \Delta} \\
\emptyset \xrightarrow{\Pi} \mathbb{V}_A \xrightarrow{\Sigma} & \mathbb{V}_{A,A} & \\
\searrow{\scriptstyle \Pi} \downarrow{\scriptstyle \Pi} & & \downarrow{\scriptstyle \Pi} \\
& \mathbb{P}_A \xrightarrow[\Sigma]{} \mathbb{P}_{A,A}. &
\end{array}$$

The bottom left triangle commutes because underlying functors do; the top left triangle commutes by Lemma A.7; the bottom square commutes by locality; the top one by Lemma A.9. □

## B.7 Prefix-based innocence

This section is devoted to the proof of Proposition 4.16. We proceed as in the previous section: we first need to show that the following diagram commutes up to isomorphism

$$\begin{array}{ccccccc}
\mathbb{P}_{A,B} + \mathbb{P}_{B,C} & \xrightarrow{\Pi} & \mathbb{P}_{(A,B),(B,C)} & \xleftarrow{\Delta} & \mathbb{P}_{A,B,C} & \xrightarrow{\Sigma} & \mathbb{P}_{A,C} \\
\downarrow{\scriptstyle \Delta} & (I) & \downarrow{\scriptstyle \Delta} & & \downarrow{\scriptstyle \Delta} & (II) & \downarrow{\scriptstyle \Delta} \\
\mathbb{P}^+_{A,B} + \mathbb{P}^+_{B,C} & \xrightarrow{\Pi} & \mathbb{P}^+_{(A,B),(B,C)} & \xleftarrow{\Delta} & \mathbb{P}^+_{A,B,C} & \xrightarrow{\Sigma} & \mathbb{P}^+_{A,C} \\
\cdots & & & \text{(Theorem 4.13)} & & & \cdots \\
\mathbb{P}^+_{A,B} + \mathbb{P}^+_{B,C} & \xrightarrow[\Pi]{} & \mathbb{P}^+_{(A,B),(B,C)} & \xleftarrow[\Delta]{} & \mathbb{P}^+_{A,B,C} & \xrightarrow[\Sigma]{} & \mathbb{P}^+_{A,C} \\
\uparrow{\scriptstyle \Delta} & & \uparrow{\scriptstyle \Delta} & & \uparrow{\scriptstyle \Delta} & & \uparrow{\scriptstyle \Delta} \\
\mathbb{P}_{A,B} + \mathbb{P}_{B,C} & \xrightarrow[\Pi]{} & \mathbb{P}_{(A,B),(B,C)} & \xleftarrow[\Delta]{} & \mathbb{P}_{A,B,C} & \xrightarrow[\Sigma]{} & \mathbb{P}_{A,C}.
\end{array}$$

(11)

By Theorem 4.13, this reduces to exactness of (I) and (II). The latter follows by hypothesis (discreteness of $\mathbb{P}^+_{A,B,C} \to \mathbb{P}^+_{A,C}$ and (II) being a pullback). The former follows by construction of $\mathbb{P}^+_{(A,B),(B,C)}$, using the zig-zag criterion.

Innocence of copycat presheaves follows similarly, as does coincidence of copycats with the essential image of restriction. Coincidence of innocence with essential image of restriction is obvious.